\documentclass[]{aa}  

\usepackage{graphicx}
\usepackage{txfonts}
\usepackage{lscape}
\usepackage[]{hyperref}
\hypersetup{breaklinks=true,colorlinks=true,urlcolor=blue, linkcolor=blue, citecolor=blue,bookmarksopen=true}
\newcommand{\AJ}{A$_J$}
\newcommand{\AK}{A$_K$}
\newcommand{\Msun}{M$_{\odot}$}

\newcommand{\Lsun}{L$_{\odot}$}

\newcommand{\Lstar}{L$_{\star}$}
\newcommand{\Mstar}{M$_{\star}$}
\newcommand{\Teff}{T$_{\rm eff}$} 
\newcommand{\Brg}{L(Br$_\gamma$)}
\newcommand{\Pab}{L(Pa$_\beta$)}
\newcommand{\Lacc}{L$_{acc}$}
\newcommand{\Macc}{$\dot{M}_{\rm acc}$}
\newcommand{\Mdust}{$M_{\rm dust}$}
\newcommand{\Mdisk}{$M_{\rm disk}$}

\newcommand{\simless}{\mathbin{\lower 3 pt\hbox {$\rlap{\raise 5pt\hbox{$\char'074$}}\mathchar"7218$}}} 
\newcommand{\simgreat}{\mathbin{\lower 3pt\hbox {$\rlap{\raise 5pt\hbox{$\char'076$}}\mathchar"7218$}}} 

\usepackage{color}

\begin{document}

   \title{The protoplanetary disk population in the $\rho$-Ophiuchi region L1688 and the time evolution of Class~II YSOs\thanks{Tables A1-G1 are only available in electronic form
at the CDS via anonymous ftp to cdsarc.u-strasbg.fr (130.79.128.5)
or via http://cdsweb.u-strasbg.fr/cgi-bin/qcat?J/A+A/}}


   \author{L. Testi
          \inst{1,2}
          \and
          A. Natta\inst{3}
          \and
          C.F. Manara\inst{1}
          \and
          I. de Gregorio Monsalvo\inst{4}
          \and
          G. Lodato\inst{5}
          \and
          C. Lopez\inst{6}
          \and
          K. Muzic\inst{7}
          \and
          I. Pascucci\inst{8,9}
          \and
          E. Sanchis\inst{1,10}
          \and
          A. Santamaria 
          Miranda\inst{4}
          \and
          A. Scholz\inst{11}
          \and
          M. De Simone\inst{12}
          \and
          J.P. Williams\inst{13}
          }

   \institute{
             European Southern Observatory, Karl-Schwarzschild-Strasse 2, D-85748 Garching bei München, Germany
             \and
             INAF-Osservatorio Astrofisico di Arcetri, Largo E. Fermi 5, I-50125 Firenze, Italy
             \and
             School of Cosmic Physics, Dublin Institute for Advanced Studies, 31 Fitzwilliam Place, Dublin 2, Ireland
             \and 
             European Southern Observatory, Av. Alonso de Córdova 3107, Casilla 19001, Santiago, Chile
             \and
             Dipartimento di Fisica, Università degli Studi di Milano, Via Giovanni Celoria 16, I-20133 Milano, Italy
             \and 
             Atacama Large Millimeter/Submillimeter Array, Joint ALMA Observatory, Alonso de Córdova 3107, Vitacura 763-0355, Santiago, Chile
             \and
             CENTRA/SIM, Faculdade de Ciencias de Universidade de Lisboa, Ed.  C8, Campo Grande, P-1749-016 Lisboa, Portugal
             \and
             Lunar and Planetary Laboratory, The University of Arizona, Tucson, AZ 85721, USA
             \and
             Earths in Other Solar Systems Team, NASA Nexus for Exoplanet System Science, USA
             \and
             Freie Universit\"at Berlin, Institute of Geological Sciences, Malteserstr. 74-100, 12249 Berlin, Germany
             \and
            SUPA, School of Physics \&\ Astronomy, University of St.~Andrews  North Haugh, St Andrews, KY16 9SS, United Kingdom
            \and
            Univ. Grenoble Alpes, CNRS, IPAG, 38000 Grenoble, France
             \and
             Institute for Astronomy, University of Hawaii, Honolulu, HI 96822, USA
             }

   \date{\today}

 \abstract
 {Planets form during the first few Myr of the evolution of the star-disk system, possibly before the end of the embedded phase. The properties of  very young disks and their subsequent evolution reflect the presence and properties of their planetary content.}
 {We present a study of the Class II/ F disk population in L1688, the densest and youngest region of star formation in Ophiuchus. We also compare it to other well-known  nearby regions of different ages, namely Lupus, Chamaeleon I, Corona Australis, Taurus and Upper Scorpius.}
 {We selected our L1688 sample using a combination of criteria (available ALMA data, {\it Gaia} membership, and optical and near-IR spectroscopy) to determine the stellar and disk properties, specifically stellar mass (\Mstar), average population age, mass accretion rate (\Macc) and disk dust mass (\Mdust). We applied the same procedure in a consistent manner to the other regions.}
 {In L1688 the relations between \Macc\ and \Mstar, \Mdust\ and \Mstar, and \Macc\ and \Mdust\ have a roughly linear trend with slopes 1.8--1.9 for the first two relations and $\sim 1$ for the third, which is similar to what found in the other regions.  When ordered according to the characteristic age of each region, which ranging from $\sim$ 0.5 to $\sim 5$ Myr, \Macc\ decreases  as $t^{-1}$, when corrected for the different stellar mass content; \Mdust\ follows roughly the same trend, ranging between 0.5 and 5 Myr,  but has an increase of a factor of $\sim$ 3 at ages of 2--3 Myr. We suggest that this could result from an earlier planet formation, followed by collisional fragmentation that temporarily replenishes the millimeter-size grain population. The dispersion of \Macc\ and \Mdust\ around the best-fitting relation with \Mstar, as well as that of \Macc\ versus \Mdust\ are equally large. 
 When adding all the regions together to increase the statistical significance, we find that the dispersions have continuous distributions with a log-normal shape and similar widths ($\sim 0.8$ dex).}
 {This detailed study of L1688 confirms  the general picture of Class II/F disk properties and extends it to a younger age. The amount of dust observed at $\sim$1Myr is not sufficient to assemble the majority of planetary systems, which suggests an earlier formation process for planetary cores. The dust mass traces to a large extent the disk gas mass evolution, even if the ratio \Mdust/\Mdisk\ at the earliest age (0.5-1 Myr) is not known.   Two properties are still not understood: the steep dependence of \Macc\ and \Mdust\ on \Mstar\ and the cause of the large dispersion in the three relations analyzed in this paper,  in particular that of the \Macc\ versus \Mdust\ relation. } 
 


   \maketitle
%

\section{Introduction}

The formation of planets in disks around young stars  has been actively studied for a long time,  especially since the discovery of exoplanets and the realization that most stars in the Galaxy are likely to host planetary systems.

Based on Solar System evidence and  the study of the incidence of disks around
young stars, it has been inferred  that  planet formation has to occur in the first few million years since the formation of the host star \citep[e.g.,][and references therein]{2014prpl.conf..547J,2015PhyS...90f8001P}. 
It is now widely accepted that planets form in protoplanetary disks, which are formed as the byproduct of the star formation process \citep[e.g.,][]{1987ARA&A..25...23S}. The lifetimes of protoplanetary disks set an upper limit to the planet formation process of about few Myr, which is consistent with the timescale for the formation of the pristine bodies in the Solar System \citep[e.g.,][and references therein]{2008ApJ...686.1195H}. 
There is increasing evidence that at least  some  planets 
form early in the evolution of the star-disk systems \citep[e.g.][]{2018A&A...617A..44K,2018ApJ...860L..13P,2019NatAs...3.1109P,2020ApJ...890L...9P,2019ApJ...883L..41C,2020ApJ...904L...6A}, most likely when the disk is still accreting matter from the collapsing core, based on the lack of solids available to form planetary cores in disks older than $0.5-1$~Myr \citep[e.g.][]{2010MNRAS.407.1981G,2014MNRAS.445.3315N,2016ApJ...828...46A,2016AA...593A.111T,2018A&A...618L...3M,2020A&A...633A.114S}
In these early evolutionary stages \citep[so-called Class~0 or Class~I protostars,][]{1987IAUS..115....1L,1993ApJ...406..122A}, the properties of the collapsing core, such as the rotation and magnetic field, are likely to affect the disk structure itself, and, as a consequence, the planet population that is undergoing formation.
The first high-angular-resolution observations by ALMA of HL~Tau, a young stellar object at the transition between Class~I and Class~II, show structures that may be interpreted as having resulted from advanced planet formation
 \citep{2015ApJ...808L...3A}.

At a slightly later stage, when the envelope has almost entirely dispersed \citep[so-called Class~II young stellar objects][]{1987IAUS..115....1L}, disks carry the imprints of the planetary systems that they host (e.g., gaps and rings), and the presence of planets affects the disk's evolution \citep[e.g.][]{Isella2016}. On the other hand, disks lose matter  due to accretion onto the central star and winds; the decrease in the gas density over time may affect, in turn, further planet growth and the dynamical evolution of  young planetary systems \citep{2016ARA&A..54..135H,2017RSOS....470114E}. The solid disk component changes over time as grains grow and drift with respect to the gas toward the star, then becoming trapped in winds and accretion flows. In addition, grains are lost to rocky planets and planetesimals, which, on the other hand, may also collide and fragment, replenishing the grain population \citep{2014prpl.conf..547J,2014prpl.conf..411T}. The environment, which hosts gravitational interactions in dense clusters and strong UV radiation field due to nearby massive stars, can also play a role \citep[e.g.,][]{2018MNRAS.478.2700W,2020Natur.586..528W}.

One possible way to study the interplay and the timeline of all these processes relies on the availability of measurements of some basic bulk quantities, such as stellar and disk masses, for a large number of objects of different masses and in different star-forming regions. The Class~II evolutionary phase lasts much longer than Class~I/0, and statistical studies of disk properties were initiated long time ago. Their results include, for example,  a measurement of the  disk lifetime of few Myr by measuring the fraction of objects with near-IR excess \citep{2008ApJ...686.1195H} and evidence of accretion \citep{2010AA...510A..72F} as function of the typical age of each region, as well as the dependence of the mass accretion rate on the age of the central star, as predicted by viscous accretion disk models \citep{2016ARA&A..54..135H}. 
Other works have provided unexpected results, such as the steep  correlation between the mass accretion rate and the stellar mass (\Macc $\propto M_{\star}^2$)  found in  Taurus \citep{2005ApJ...625..906M} and $\rho-Oph$ \citep{2006AA...452..245N}, which was interpreted as having resulted from a spread in the properties of the parental cores \citep{2006ApJ...645L..69D}, although alternative explanations have not been excluded \citep[e.g.][]{2014MNRAS.439..256E,2009ApJ...703..922V,2006ApJ...639L..83A,2006MNRAS.370L..10C,2006ApJ...648..484H,2006MNRAS.371..999G}.

It is, however, only in the last decade that with the advent of large surveys both in the optical and mm wavelengths, these kinds of studies have become possible for a large number of star-forming regions and possible evolutionary patterns have been established.
In particular, the greatly improved sensitivity provided by VLT's second-generation instruments and by ALMA have allowed for population studies of accretion and disk properties to be exptended down to very-low-mass stars (VLMS) and brown dwarfs (BDs), and, more generally, lower-mass disks, allowing us to effectively undertake broader population studies
\citep[][]{2016AA...593A.111T,2020A&A...633A.114S}.
The broad wavelength range and sensitivity of VLT/X-Shooter has  been used to derive reliable stellar parameters and mass accretion rates \citep[e.g.][]{2017AA...600A..20A,2015AA...579A..66M,2017A&A...604A.127M,2020AA...639A..58M}, while complementary ALMA surveys have allowed us to measure millimeter dust continuum and molecular line emission  for disks over a large range of stellar masses, from brown dwarfs to solar-mass stars \citep[e.g.][]{2016ApJ...828...46A,2016ApJ...831..125P,2016ApJ...827..142B,2016AA...593A.111T,2019A&A...626A..11C}. While the original hope to  constrain the disks gas masses from the observations of the carbon monoxide isotopologues have turned out to be more uncertain than originally expected \citep[e.g.,][]{Miotello2017LupusDepletion,2017ASSL..445....1B}, several complementary lines of evidence suggest that the millimeter continuum flux is a good proxy of the solids mass in disk albeit with systematic uncertainties due to the poorly constrained dust physical and chemical properties \citep[][]{2014prpl.conf..411T,2015PASP..127..961A}. Moreover, a number of recent studies have started to investigate the  different evolution of the dust millimeter grains and the gas component of the disk by including the effects of growth, drift, and fragmentation  \citep[][]{2020A&A...635A.105P,2020MNRAS.498.2845S,2021AJ....162...28V,2021arXiv210405894M}. On the other hand, the few available attempts to directly measure  disk masses from gas kinematical modeling show a remarkable consistency  with estimates from the dust continuum measurements \citep[assuming the canonical gas to dust ratio of 100, e.g.][Izquierdo  et al., in preparation]{2021ApJ...914L..27V}.

Notwithstanding the above uncertainties, several important trends for disk bulk properties have been confirmed: the steep dependence of \Macc\ on \Mstar\  \citep[e.g.][]{2017AA...600A..20A,2017A&A...604A.127M,2020AA...639A..58M}, and the steeper than linear correlation between \Mdust\ and \Mstar\ \citep[e.g.][]{2016ApJ...828...46A,2016ApJ...831..125P,2016ApJ...827..142B}. \citet{2016AA...591L...3M}  showed for the first time that \Macc\ increases almost linearly with the disk (dust) mass, as predicted by viscous evolution models. This was also confirmed by \citet{2017ApJ...847...31M}, who also pointed out that the dispersion of the data points cannot be easily reconciled in the simplified viscous evolutionary models, unless the age of the disks is comparable with the viscous timescale \citep{2017MNRAS.472.4700L}. 
A clear trend showing a decrease in \Mdisk (or \Mdust) with the age of the region was found when comparing the 2-3~Myr Lupus and Chamaeleon~I with the 5-8 Myr Upper Scorpius \citep[][]{2017AJ....153..240A}.
Other intriguing results have also been reported. For example, Upper Scorpius has large mass accretion rates and a huge dispersion of values when compared to Lupus and Chamaeleon~I for disks of the same mass \citep{2020AA...639A..58M}, contrary to the expectations of viscous evolutionary models \citep[see e.g.][]{2017MNRAS.472.4700L,2017MNRAS.468.1631R}.

Very recently, the ALMA data of two very young regions, $\rho-Oph$ \citep[]{2019ApJ...875L...9W,2019MNRAS.482..698C}  and Corona Australis \citep[]{2019A&A...626A..11C} seem to indicate that the disk (dust) mass in these regions is smaller than in the older Lupus and Chamaeleon~I regions. This is unexpected, and potentially very interesting, if confirmed. However, the results for $\rho-Oph$ could be contaminated by a significant component of older objects, belonging to groups of different ages, including the overlapping Upper Scorpius population, in the same region \citep[][]{2020AJ....159..282E}; moreover, only the cumulative mass distribution of the regions were compared, as a characterization of the stellar properties of the objects (\Mstar\ in particular) was not available.

In this paper we present a new analysis of the young stellar photosphere and disk properties for the Class~II/F protoplanetary disks in the $\rho$-Ophiuchi L1688 region. 
This sample is then compared with other well known star forming regions in the Solar neighborhood, namely Lupus, Chamaeleon~I, Corona Australis, Taurus and Upper Scorpius, with the aim of analyzing potential evolutionary trends. In all  cases, we base our analysis on new membership and distance information derived from spectroscopy and the Gaia DR2  data, and we derive the disk and stellar properties in a homogeneous way.

In Sect.~\ref{oph_samp} and~\ref{sec:res_l1688}, we describe the $\rho$-Ophiuchi L1688 sample properties, new observations, and results. In Sect.~\ref{other_samp}, we describe the properties of the objects in the other star-forming regions analyzed in this paper.  The results are discussed in Sect.~\ref{sec:disc}. Conclusions follow in Sec.~\ref{concl}.

\section {Ophiuchus L1688 sample}
\label{oph_samp}

We focus our study on L1688, the densest and youngest region of star formation in Ophiuchus \citep{2008hsf2.book..351W,2020AJ....159..282E}. 
We selected young stars with disks classified as Class II or F,
observed with ALMA, and with estimates of three crucial photospheric parameters, namely spectral type (SpT), extinction (\AK), and $J$-band magnitude, which allow us to estimate the stellar properties. In the following sections we summarize the details of the selection process.

\subsection {ALMA L1688 data}

\subsubsection{ALMA 1.3mm data from the ODISEA survey}
The Ophiuchus region was the subject of an extensive 1.3mm ALMA survey of the young stellar population  \cite[The Ophiuchus DIsk Survey Employing ALMA (ODISEA),][W19 in the following]{2019MNRAS.482..698C,2019ApJ...875L...9W}. The survey includes all the   protostars identified in the {\it Spitzer} “Cores to Disks” Legacy project \citep{2009ApJS..181..321E} for a total of 279 objects  (Class I, Flat spectrum, Class II and Class III).  Of these, 172 are classified as Class~II and 50 as Class~F\footnote{In this paper we adopt the existing SED classifications based on {\it Spitzer} observations, as defined in the cited references for each region}. The surveyed area covers, but extends beyond L1688,  including, for example, members of the older Upper Scorpius association and of other young groups in the $\rho$--Ophiuchi star forming region (e.g. L1689 and L1709, among others). We will discuss in 
Sect.~\ref{sec:EL20} how we selected from the ODISEA survey the members of L1688 that will be used in this study.

\subsubsection{ALMA 0.89mm data for VLMS and BDs}
\label{sec_alma}

As part of a survey of very low mass stars and brown dwarfs with disks, we observed with ALMA a total of 25 objects at 0.89mm with ALMA \citep[T16 from now on, and this work]{2016AA...593A.111T}. A detailed discussion of this sample is given in Appendix~\ref{sec:Alma-BDs}.


\subsection {Membership and spectroscopy}
\label{sec:EL20}

\citet[][EL20 in the following]{2020AJ....159..282E} have recently published a new, critical analysis of the Ophiuchus region. Their  aim was to   establish the membership of all known objects with evidence of youth in addition to known spectral type  in a region which includes the three main dark clouds of Ophiuchus (L1688, L1689, L1709) and a substantial area north of L1688, encompassing the majority of W19 ALMA region. They separated the Ophiuchus subgroups members from other populations by applying kinematic criteria to the Gaia second data release. The result is a catalog of 373 members, 259 candidate members and 59 stars with kinematics that are inconsistent with membership in Ophiuchus. 

We focus on L1688, the youngest core of star formation in the region. It occupies an area on the plane of the sky defined 
according to Fig.1 of EL20.  In this region, there are 154 young stellar objects classified as Class~F or Class~II by W19 and \citet[][based on previous Spitzer surveys]{2019MNRAS.482..698C}.
Of these, 88 are classified as members of L1688 and 7 as candidate members by EL20. For all these objects we adopted the spectral type, extinction and J-band magnitude as given by EL20.

To these we added three sources (J162435.2-242620, J162618.1-243033, J162732.6-243323) in the L1688 region, for which EL20 does not provide spectral types. However, we did found  spectroscopically determined spectral types, extinction, and J-band magnitude values for them in the literature \citep{2016AA...592A.126V,2012ApJ...744..134M,2010ApJS..188...75M}. 

Seven of the BDs observed in Band~7 by T16 or in this study are not present in the ODISEA survey (GY92-320 from T16, and GY92-141, CFHTWIR-Oph-58, CFHTWIR-Oph-66, CFHTWIR-Oph-77, CFHTWIR-Oph-90, CFHTWIR-Oph-98, from this paper). They are all classified as member or candidate member by EL20, and we adopt their parameters from EL20, as above.

In summary, the L1688 sample analyzed in this work contains 105~objects. 

\subsection {Stellar Parameters}

Using the J-band magnitude, the extinction and the spectral type, for each source we estimated photospheric luminosity \Lstar\ and effective temperature \Teff. 
We derived the J-band extinction from \AK\ using the ratio \AJ/\AK=2.63 (EL20) and computed for each star the effective temperature \Teff\ according to the SpT-\Teff\ relation used by  \citet{1998AAS..133...81T} and \citet{2003ApJ...593.1093L} \citep[see also the references quoted in these papers, as well as the analysis in][]{2009AA...503..639T}.
The luminosity is then derived from the J-band magnitude, corrected for extinction, the { Gaia} distance, and the bolometric correction to the J magnitude as compiled by \citet{1998AAS..133...81T} and \citet{1999ApJ...525..466L}. For the stars with no { Gaia} distance we adopt the average value of 139.4~pc (EL20).

We proceeded in the same way for  objects with extinction and spectral type from other studies. In each case, we used the extinction law quoted in the relevant paper to convert to a consistent set of J-band extinctions (see Sect.~\ref{sec:EL20}). 

\begin{figure*}
   \centering
       \includegraphics[width=18cm]{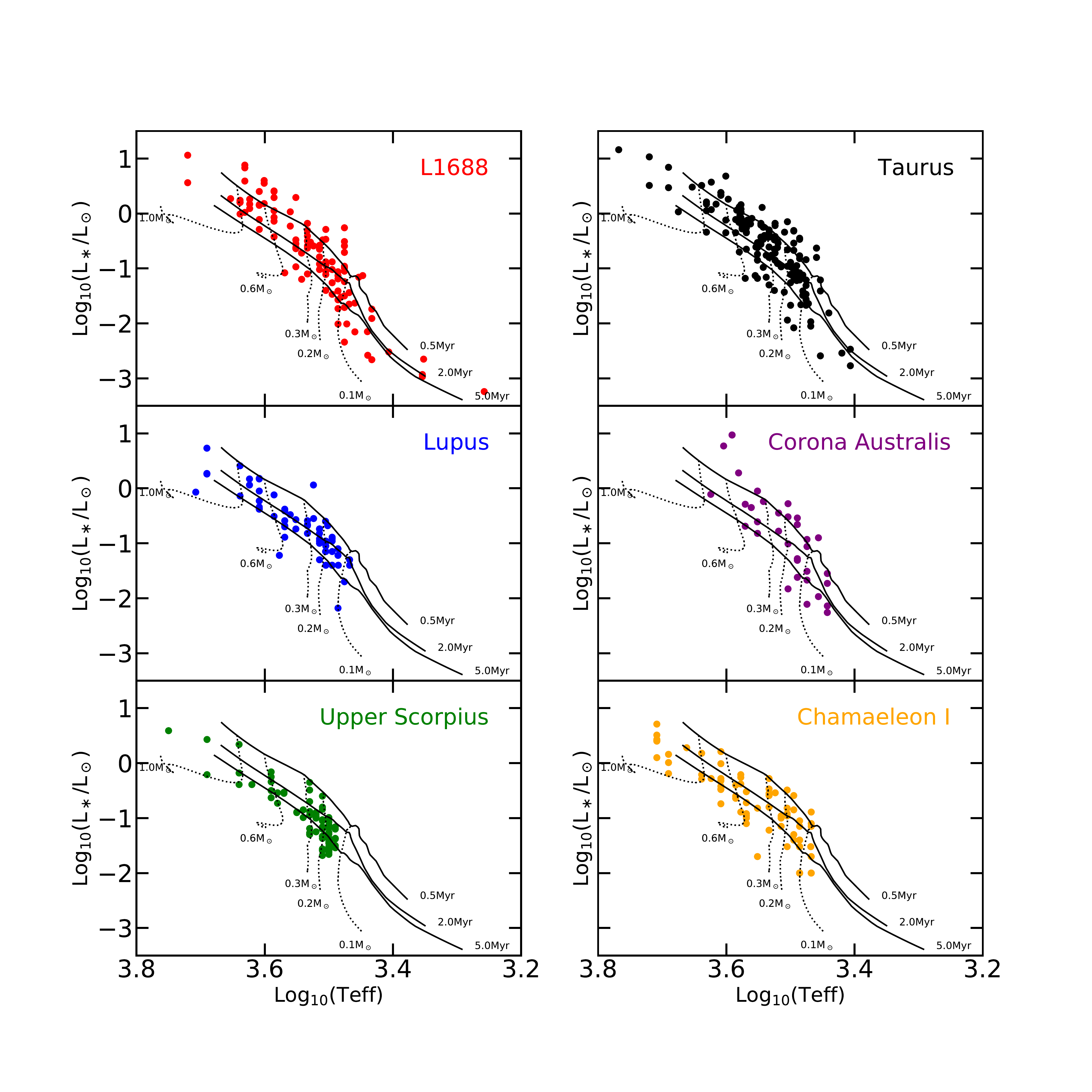}
      \caption{HR diagrams for the six star-forming regions, as labelled in each panel. 
      Stellar evolutionary tracks (dotted lines) and isochrones (solid lines) are from \citet{2015AA...577A..42B}; masses and ages are as labelled.
              }
         \label{Fig:HR}
   \end{figure*}

The resulting Hertzsprung-Russel (HR) diagram is shown in Fig.~\ref{Fig:HR} (top left panel), with overlaid the tracks from \citet{B15}.
Uncertainties on both \Teff\ and \Lstar\ can be very large,  especially in a region of high extinction as L1688 (see, e.g., EL20). \citet{2015AA...579A..66M} quote at least one spectral sub-class and $>0.2$dex in \Lstar, even when using X-Shooter spectra with their broad wavelength coverage and high sensitivity. As discussed in the literature \citep[e.g.,][]{2017AA...600A..20A}, uncertainties in the spectral classification may be larger for G- and K-types, as compared to M-dwarfs. We thus estimated an error of 1.5--2 sub-classes for the higher-mass objects in our sample.

Stellar masses are derived by comparing the location on the HR diagram to the pre-main sequence evolutionary tracks of \citet{B15}, when possible; for objects with stellar mass larger than 1.4 \Msun, we used the \citet{Siess00} tracks. The choice of these evolutionary tracks is meant to facilitate the comparison with previous works on a number of star-forming regions \citep[e.g.][]{2017AA...600A..20A,2020A&A...633A.114S}. 
For a given set of evolutionary tracks, and for the bulk of the stellar masses and ages covered in our sample, the evolutionary tracks are nearly vertical in the HR diagram; hence, the derived values of \Mstar\ are affected mostly by  uncertainties in the spectral type. Values of \Mstar\ for objects more luminous than the evolutionary tracks was derived  by vertically extrapolating the tracks.
Typical uncertainties  are at least $\pm 10$\%, but can be significantly larger for individual objects with highly uncertain spectral type.

%

\begin{figure*}
   \centering
       \includegraphics[width=18cm]{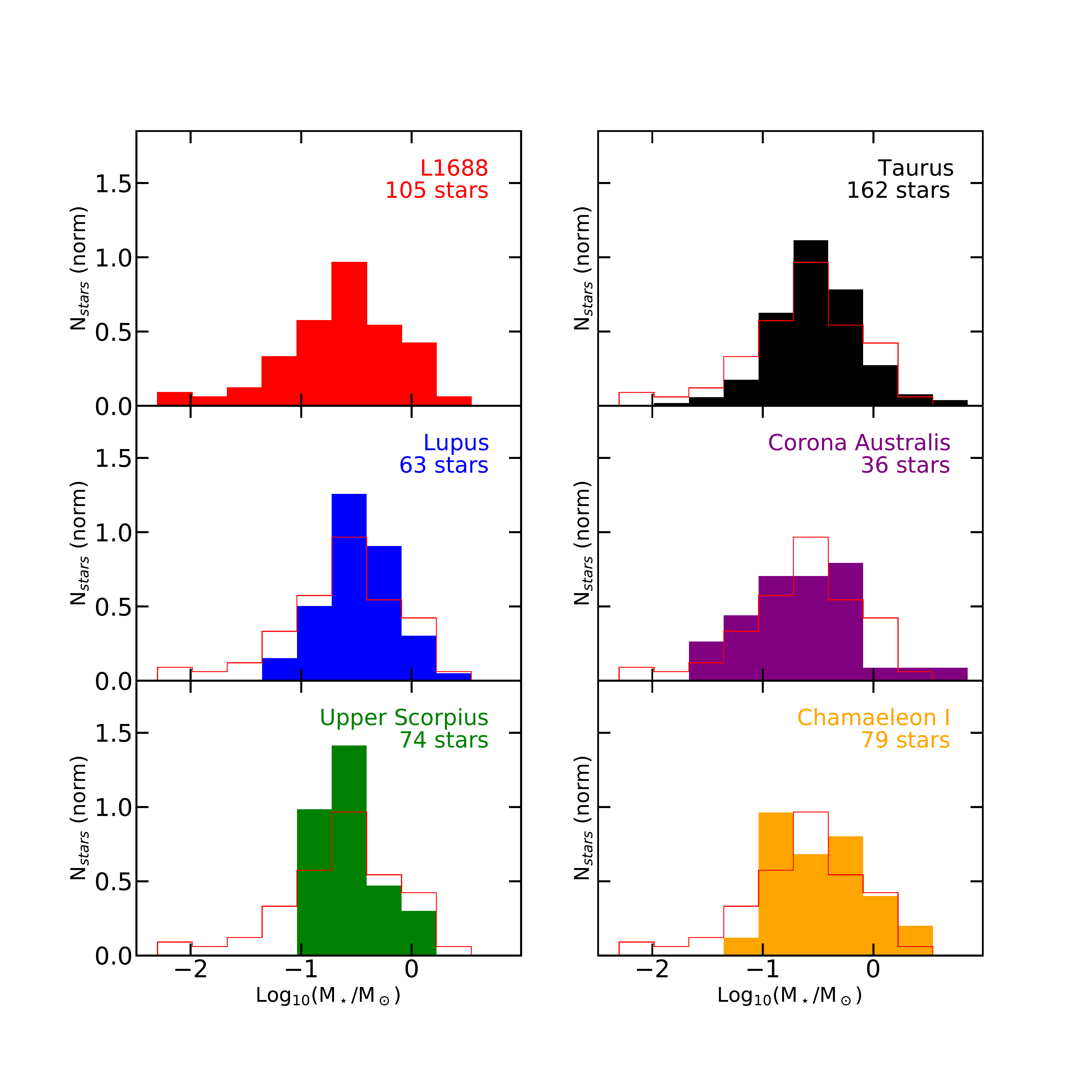}
      \caption{Stellar mass distribution for the six regions, as labelled. The total number of stars in each regions is given in the respective panel. The distribution for L1688 is shown in all the other panels by the red line.
              }
         \label{Fig:Mstar_hist}
   \end{figure*}

\subsection{Mass accretion rate}
\label{Sec:Macc_L1688}

The largest survey of mass accretion rates in  L1688 was published by \citet{2006AA...452..245N}.  Accretion luminosities were computed for a sample of 104 Class II/F stars using the \Pab\ in 96 cases or \Brg\ for the remaining 8 luminosity as a proxy for \Lacc.  Of the 105 objects in our sample, 60 have been observed by \citet{2006AA...452..245N}. 
We recomputed the line luminosity for the distance and extinction adopted in this paper and derive \Lacc\ from the relations of \citet{2017AA...600A..20A}. 

We then derived the mass accretion rates  from the relation $\dot M_{acc}= 1.25 L_{acc} R_\star/(G M_\star)$, with $R_\star$ computed from the photospheric luminosities and effective temperatures.

We note that in \citet{2006AA...452..245N}, who selected their sample from the ISOCAM YSOs in the study of \citet{2001AA...372..173B},  a total of 140 objects were observed, of which 35 were classified as Class~III and 105 as Class~II. The 45 ISOCAM Class~II objects excluded from the sample discussed in this paper are either not present in the ALMA ODISEA survey or are classified as non-members of L1688 by EL20. 

\section{Results for L1688}
\label{sec:res_l1688}

The L1688 sample of 105 objects is the largest  among the nearby star forming regions (d$< 150-160$~pc), spanning a large range of spectral types, stellar luminosities (between $\sim 10^{-3}$ and 10~\Lsun), and masses (from $\sim$0.01 to $\sim$1~\Msun).  The \Mstar\ distribution  is shown in Fig.~\ref{Fig:Mstar_hist}.


\subsection{Relationship between disk dust mass and stellar mass}
\label{Sec:l1688_mdisk}

The dust mass in each disk is computed from the millimeter flux, assuming optically thin emission and Rayleigh-Jeans approximation, according to:
\begin{equation}
    M_{dust}= {{F_{mm} \,d^2} \frac {\kappa _\nu}{B_\nu(T_{dust})} }
\end{equation}
where $F_{mm}$ is the observed flux, $d$ the distance, $\kappa _\nu$ the dust opacity at the frequency of the observations, and $B_\nu$ the Planck function at the dust temperature $T_{dust}$. Adopting $\kappa _\nu$=2.3 cm$^2$/g at 1.3mm and $T_{dust}=20$K, as in \citet{2017AJ....153..240A}, Eq.(1) gives $M_{dust}/M_\odot\sim 9 \times 10^{-11} d^2 F_{1.3mm}$ with $d$ in pc and $F_{1.3mm}$ in mJy.
For the 7 L1688 BDs with only 0.89mm fluxes from ALMA, we assume a ratio of $F_{1.3mm}/F_{0.89mm}$=0.44 (see Appendix~\ref{sec:Alma-BDs}). 

In the following, we define the total (gas + dust) disk mass as \Mdisk= $100 \times$\Mdust. Also, we  use  \Mdust\ or \Mdisk\ depending on the context. We note that, in fact,  both are simply proportional to the measured millimeter flux.

\begin{figure}
   \centering
       \includegraphics[width=8.8cm]{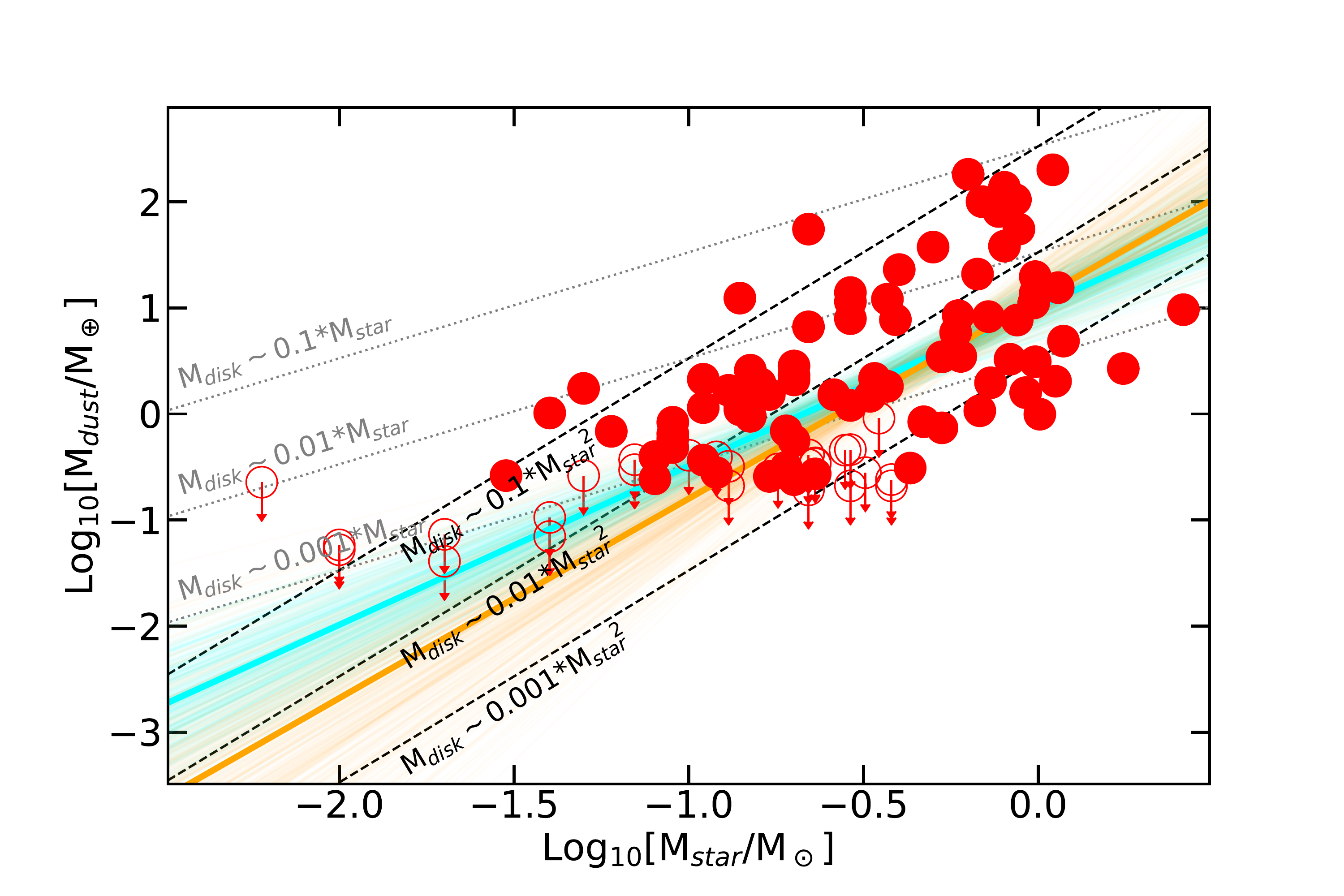}
      \caption{
      Dust mass, \Mdust,   as a function of the stellar mass \Mstar\  for L1688. Filled dots are actual measurements, open dots with arrows  are  upper limits. Cyan lines plot the results of a linear fit over the whole range of \Mstar; orange lines the results when stars with \Mstar<0.15 \Msun\ are excluded (see text).
      For comparison, we plot the relations  \Mdisk $\propto$ \Mstar$^2$ (dashed black lines) and \Mdisk $\propto$ \Mstar (dotted grey lines) for different values of the  \Mdisk/\Mstar\ ratio, as labelled. Note that, as always, \Mdisk=100 $\times$ \Mdust.
              }
         \label{Fig:Mdust-Mstar_Oph}
   \end{figure}

The distribution of \Mdust\ as a function of \Mstar\ is shown in Fig.\ref{Fig:Mdust-Mstar_Oph}. 
Compared to the compilation of \citet{2016AA...593A.111T}, this is a significant improvement.  With the exception of a handful of BDs, all disk masses come from a coherent set of ALMA  measurements at 1.3mm; the membership and stellar parameters have been derived in a consistent and more accurate way for a sample that is three times greater, covering  the range of stellar masses in a more uniform way. 

We note that for stars above $\sim$0.4M$_\odot,$ all disks have been detected at 1.3mm, thanks to the sensitivity of the ODISEA ALMA survey \citep{2019MNRAS.482..698C}. The ~two orders of magnitude in dispersion for  \Mdust\ at any given \Mstar\ is consistent with what is observed in other nearby star forming regions \citep[e.g.,][]{2016ApJ...831..125P,2017AJ....153..240A,2016ApJ...828...46A}.

Figure~\ref{Fig:Mdust-Mstar_Oph}  shows   lines  \Mdisk $\propto$ (\Mstar/\Msun) and \Mdisk $\propto$ (\Mstar/\Msun)$^2$, for different proportionality coefficients, as labeled.
We confirm the results presented by \citet{2016AA...593A.111T} in their study of  the dust masses in disks around BDs , namely that the relation  between \Mdust\ and \Mstar\ when a sufficiently broad range of \Mstar\ is considered  is steeper than linear. This result has since been confirmed by   most nearby Class~II disk populations studied so far  \citep[e.g.][]{2016ApJ...831..125P,2017AJ....153..240A}.
To quantify the relationship between \Mdust\ and \Mstar, following the approach used in similar studies, we performed a power law fit to the data shown in Fig.~\ref{Fig:Mdust-Mstar_Oph}, using the method developed by \citet{2007ApJ...665.1489K} (see Appendix~\ref{App:fits} for details). 
When using the full sample, including upper limits, we find
that \Mdust$\sim$\Mstar$^{1.5}$ (cyan thick line in figure), but we note that this result is mostly constrained by the limited data in the BDs regime. A fit   for \Mstar$\ge 0.15$\Msun\ gives    \Mdust$\sim$\Mstar$^{1.9}$ (orange thick line in figure), with an uncertainty of $\sim$0.4. 

\subsection{Relationship between mass accretion rate and stellar mass}

\begin{figure}
   \centering
       \includegraphics[width=8.8cm]{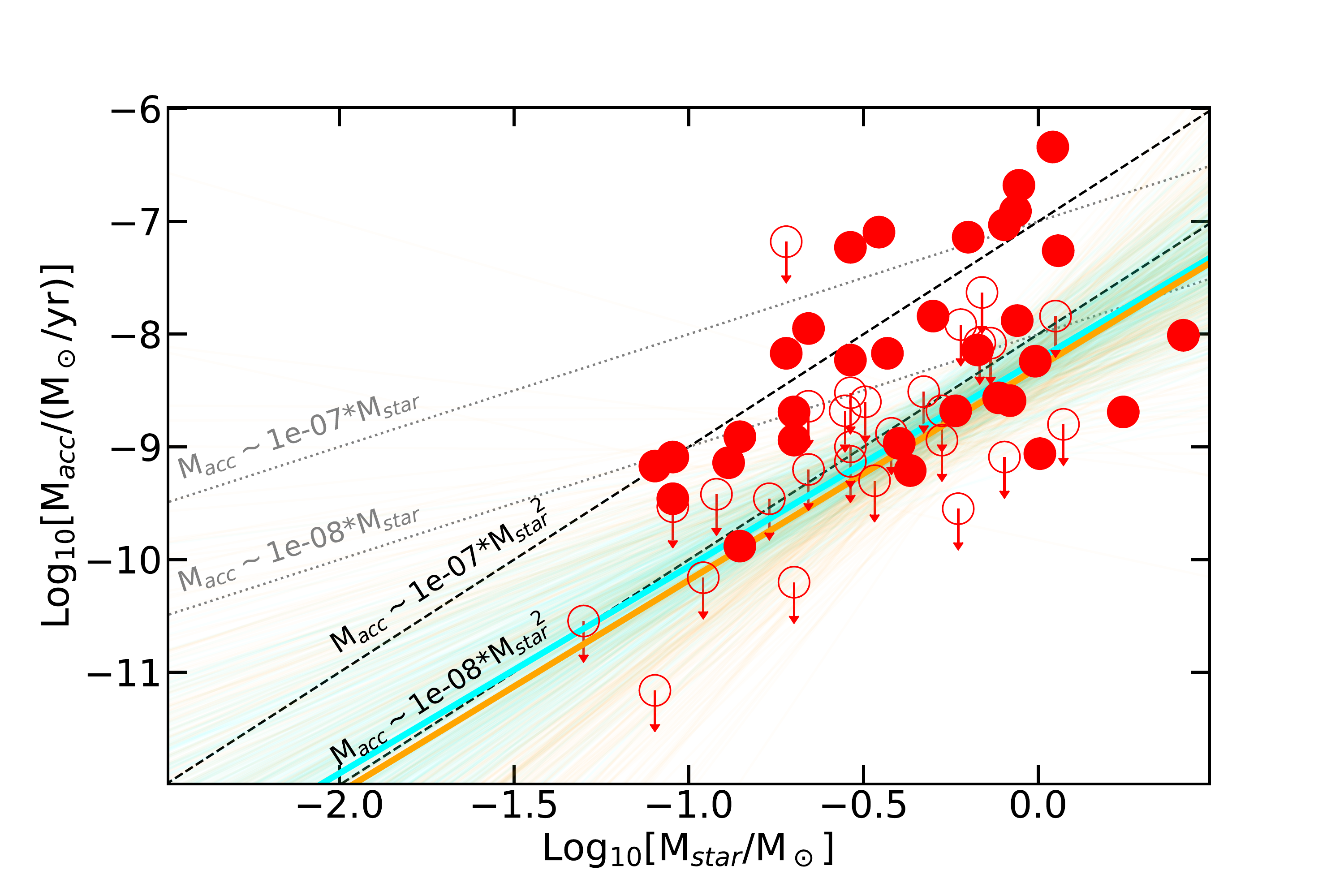}
      \caption{ Mass accretion rate  as a function of \Mstar\ for L1688; symbols as in Fig.~\ref{Fig:Mdust-Mstar_Oph}. Cyan lines show the results of a linear fit over the whole range of \Mstar; orange lines the results when stars with \Mstar<0.15 \Msun\ are excluded (see text).
      For comparison, we plot the relations  \Macc $\propto$ \Mstar$^2$ (dashed black lines) and \Macc $\propto$ \Mstar (dotted grey lines) for different values of the  \Macc/\Mstar\ ratio, as labeled.
              }
         \label{Fig:Macc-Mstar_Oph}
   \end{figure}

Figure~\ref{Fig:Macc-Mstar_Oph}  shows the measured \Macc\ values and upper limits in L1688 as a function of the stellar mass. Grey dotted and black dashed lines  show the relations expected for a linear and quadratic dependence of the mass accretion rates on the stellar mass. As already noted in other star-forming regions  \citep[e.g.][]{2005ApJ...625..906M,2006AA...452..245N,2017AA...600A..20A,2017A&A...604A.127M}, when extending to the very low mass stars and BDs, the \Macc\ values are lower than expected from a linear relation. 

To quantify the relationship between \Macc\ and \Mstar, following the approach used in similar studies, we performed a power law fit to the data shown in Fig.~\ref{Fig:Macc-Mstar_Oph}, using the method developed by \citet{2007ApJ...665.1489K} (see Appendix~\ref{App:fits} for details). 
When using the full sample, including upper limits, we find
that \Macc$\sim$\Mstar$^{1.9}$ (cyan thick line in figure). A fit   for \Mstar$\ge 0.15$\Msun\ gives a similar result, within the uncertainties (\Macc$\sim$\Mstar$^{1.8}$; orange thick line in figure). 
These values are nearly identical to the results of \citet{2006AA...452..245N}, who did not have a significant number of spectral type classifications for the Ophiuchus Class~II YSOs; hence, their stellar masses were derived from the extinction corrected J-band luminosities assuming a single 1~Myr isochrone.  The net result is that photospheric parameters derived in our previous study were more uncertain and that the upper limits were correlated with the stellar luminosity and mass, as they were  also derived from the  J-band luminosities \citep[see][for details of the procedure]{2006AA...452..245N}. 


\section{Comparison with other star-forming regions}
\label{other_samp}

The new ALMA data, membership determination, and stellar properties set a much firmer foundation for the characterization of L1688 disk properties, allowing us to add it  to the sample of the other nearby star-forming regions to study   the evolution of disk populations. For this purpose, we selected a sample of the best-studied nearby regions and re-determined their stellar and disk properties, as described in this section,  to ensure an   approach that is as homogeneous as possible.

\subsection {Data selection}
\subsubsection{Lupus}
\label{lup_samp}

The Lupus disk population has been extensively studied with ALMA by \citet{2016ApJ...828...46A,2018ApJ...859...21A,2020A&A...633A.114S}, and the photopheric and disk accretion properties with XShooter by \citet{2014AA...561A...2A,2017AA...600A..20A}. Recently, \citet{2020AJ....160..186L}  re-assessed the membership to the Lupus cloud of all candidates using Gaia data, as in EL20. We apply the same criteria described in Sect.~\ref{sec:EL20} to the selection of objects of Class F/II that are members or candidate members according to \citet{2020AJ....160..186L}, and have been observed  with ALMA. This results in a reduction from the original 95 objects observed by \citet{2018ApJ...859...21A} to the 63 studied in this paper. All these objects have a determination of stellar parameters and accretion rates from X-Shooter observations \citep{2017AA...600A..20A}, which we recomputed using the new Gaia distances. Disk masses are computed from the data in the 
1.3mm survey of \citet{2018ApJ...859...21A}, with the same prescription used for L1688.

The table of the adopted properties for the Lupus sources is reported in Appendix~\ref{App:tab_lupus}. 

\subsubsection{Upper Scorpius}
\label{usc_samp}

 A fraction of the Upper Scorpius disk population has been surveyed with ALMA at 0.88 mm by \citet{2016ApJ...827..142B}, who also report the photospheric effective temperatures and luminosities for their sample. While it remains incomplete, this is nonetheless the largest homogeneous millimeter survey of Upper Scorpius available to date. \citet{2020AJ....160...44L}  analyzed the membership properties of Upper Scorpius using Gaia data. We applied the same selection criteria as before to define the sample of the 74 Upper Scorpius Class~F and~II discussed in the following, using, as before, the new Gaia distances to recompute photospheric and accretion properties. \citet{2020AA...639A..58M}  sampled a subset of 34~sources from the \citet{2016ApJ...827..142B} study with XShooter to derive disk accretion properties -- all of these are included in our sample.  We note that \citet {2020AA...639A..58M} has already used the new Gaia distances. Disk masses are computed from the \citet{2016ApJ...827..142B} fluxes converted to 1.3mm using the average ratio 0.4 (spectral index 2.4) as derived for the disk populations of several nearby star-forming regions
\citep{2018ApJ...859...21A,2017ApJ...849...63R,2020arXiv201002249T}, and Eq.(1). The table of the adopted properties for the Upper Scorpius sources is reported in Appendix~\ref{App:tab_usc}. 

\subsubsection{Chamaeleon~I}
\label{cha_samp}

A large fraction of the Chamaeleon~I disk population has been surveyed with ALMA at 0.89 mm by \citet{2016ApJ...831..125P}, while \citet{2017A&A...604A.127M} carried out an 
extensive spectroscopic campaign to accurately determine stellar photospheric properties and accretion rates.
Following \citet{2019A&A...631L...2M}, we used Gaia DR2 to correct the derived parameters using the new individual (for the stars in the Gaia catalogue) and average distances. 
Starting from  \citet{2016ApJ...831..125P}, and applying the same selection criteria as before we derive a sample of 79 Chamaeleon~I Class~F and~II stars, which we use in the following. 
The Gaia membership of the Cha I region was recently  analyzed by \citet{2021A&A...646A..46G}, and 64 of our 79 stars are confirmed members, with their distance estimates  consistent with our values. Of the remaining 15 stars, 13 do not pass the strict quality standards of \citet{2021A&A...646A..46G}, but are classified as members by \citet{2017AJ....154...46E}; the remaining two stars are considered candidate members according to the criteria of \citet{2016ApJ...831..125P} and \citet{2017A&A...604A.127M}.
%
%
As for Upper Scorpius, we computed disk masses from the 0.89mm fluxes converted to 1.3 mm fluxes with the same constant factor of 0.4 and eq.(1).
The table of the adopted properties for the Chamaeleon~I sources is reported in Appendix~\ref{App:tab_cha}.

\subsubsection{Corona Australis}
\label{cra_samp}

The Corona Australis regions was surveyed with ALMA by \citet{2019A&A...626A..11C}, who also reports on the distance estimate and the photospheric parameters of the stars. The Gaia membership of the Corona Australis region was recently  analyzed by \citet{2020A&A...634A..98G}, and only 15 of our 36 stars are present in the Gaia members catalog, probably because of the strict quality threshold applied on the data and the high extinction in the Corona Australis region. We thus adopted as candidate members all the sources in  \citet{2019A&A...626A..11C}.
The table of the adopted properties for the Corona Australis sources is reported in Appendix~\ref{App:tab_cra}. 

\subsubsection{Taurus}
\label{tau_samp}

The most complete compilation of submillimeter continuum measurements of young stars with disks in the Taurus region is  that of \citet{2019ApJ...872..158A}. Recently \citet{2019AJ....158...54E}  revisited the membership and properties of the Taurus young stars using Gaia data. To derive the stellar photospheric parameters we thus followed the same procedure as in the L1688 region (see Sect~\ref{sec:EL20}). 
The table of the adopted properties for the Taurus sources is reported in Appendix~\ref{App:tab_tau}.  

\subsection{Stellar mass distributions  and ages}

Figure~\ref{Fig:HR} shows the HR diagram for the six regions, while the mass distribution is shown in Fig.~\ref{Fig:Mstar_hist}. While only L1688 -- and to some extent Taurus -- extends to very low mass objects and BDs, above \Mstar$\sim 0.15$~\Msun,\ all the regions are well sampled. 

A comparison of the HR diagrams shows that L1688 is the youngest, with ages similar to those of Taurus and Corona Australis; Lupus and Chamaeleon are likely older by $\sim$1--2 Myr, while Upper Scorpius is about 5 Myr.  This agrees with the relative age distribution in the literature, where   
there is general consensus  that L1688 is one of the youngest region of star formation, several Myr younger than Upper Scorpius \citep[see, e.g.,][]{2020AJ....159..282E}. 
For each region, Table ~\ref{tab:Region_Ages} shows the median values, first and third quartiles of the stellar age distribution of the stars in our samples, derived from comparison to the  \citet{2015AA...577A..42B} evolutionary tracks. To avoid the parameter range where the evolutionary tracks are closer to each other than the uncertainties on the stellar temperature and luminosity,  we considered only stars with masses in the range of 0.15$\le$M$_\star$/M$_\odot\le$1.0.
In the following, we use these as the representative ages for our samples of Class~II objects in each region. The uncertainties on age estimates derived by comparing the location of the stars on the HR diagram with evolutionary tracks are well known  \citep[see, e.g.,][]{2017MmSAI..88..574D,2015ApJ...808...23H,2021ApJ...908...46B} and beyond the purpose of this paper. 

\begin{table}
    \centering
    \begin{tabular}{c c c c} 
    \hline
    Name& Median Age& 25\%& 75\%\\
    & (Myr)& (Myr) & (Myr)\\
    \hline
      Corona Australis  & 0.6 & 0.5 & 2.1\\
      L1688    &  1.0 & 0.5 & 2\\
       Taurus & 0.9& 0.5& 1.7\\
       Lupus & 2.0 & 1.3& 3.6\\
       Chamaeleon I & 2.8& 1.4& 6.6\\
       Upper Scorpius& 4.3& 2.7& 7.6\\
       \hline
    \end{tabular}

    \caption{Characteristic age of the regions. Median, lower and upper quartiles of the age of the stars in each region, derived from a comparison with the \citet{2015AA...577A..42B} evolutionary tracks. In computing the ages, we considered only stars with masses in the range 0.15$\le$M$_{star}$/M$_\odot\le$1.0.}
    \label{tab:Region_Ages}
\end{table}

\subsection{\Mdust\ cumulative distributions}
\label{Sec:MdustCum}

One methodology that has been widely used in the past to compare the typical disk dust mass in different star-forming regions is the cumulative mass distribution \citep[e.g.,][]{2013ApJ...771..129A,2016ApJ...828...46A}.

\begin{figure}
   \centering
       \includegraphics[width=8.8cm]{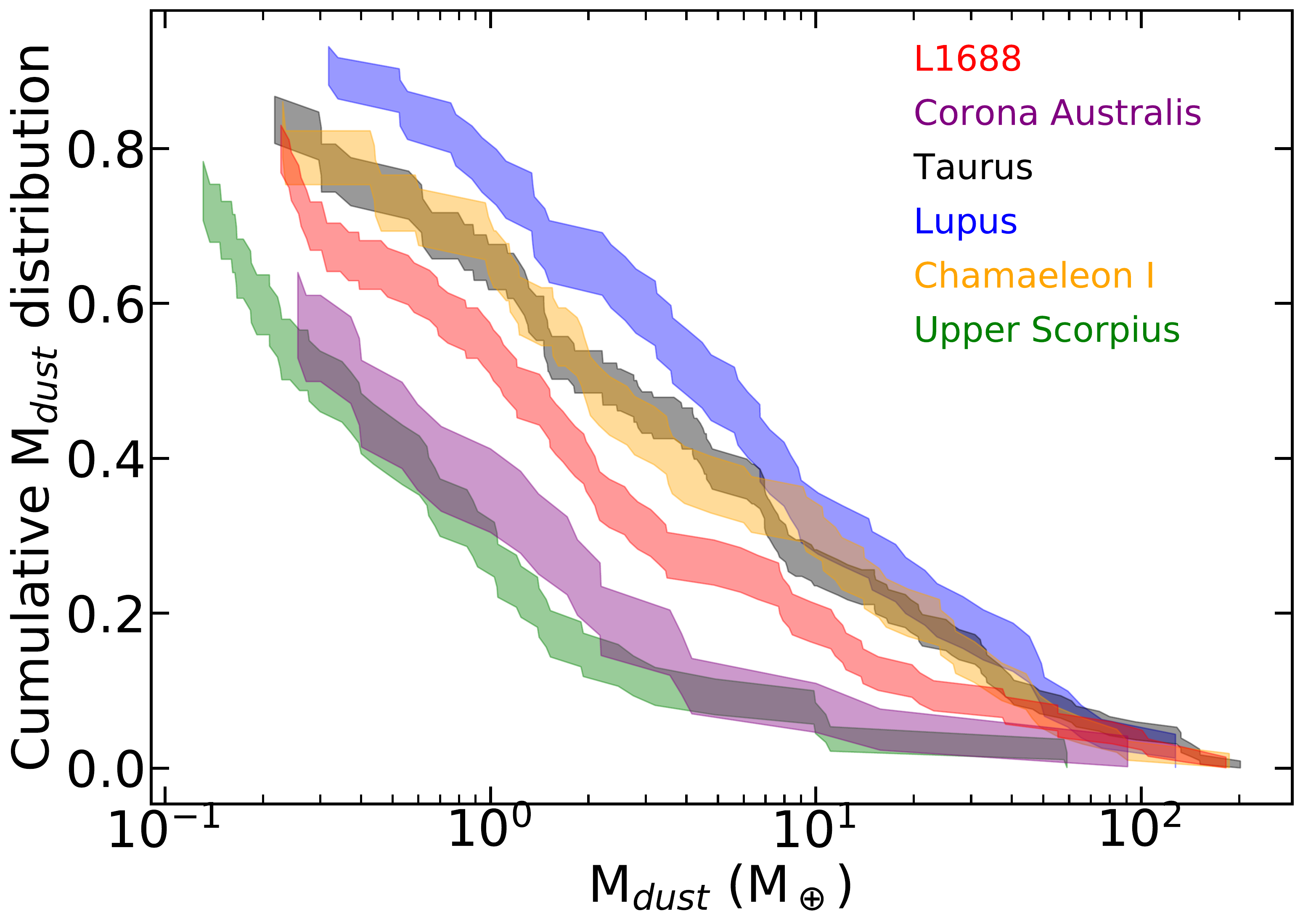}
      \caption{Comparison of the \Mdust\ cumulative distributions in the various regions. The color-shaded regions show the uncertainty range for each distribution. 
              }
        \label{Fig:Mdust_Cumul}
\end{figure}

Figure~\ref{Fig:Mdust_Cumul} shows the dust mass cumulative distribution for each star-forming region computed using the Kaplan-Meier estimator from the \texttt{lifelines}\footnote{\url{https://lifelines.readthedocs.io}} package, which takes into account the upper limits using a proper technique for left censored datasets. 
Lupus, Taurus, and Chameleon have a similar cumulative function for the disk (dust) mass and differ from those of Upper Scorpius and Corona Australis, while the L1688 cumulative dust mass function is placed between these two groups.
This confirms the results already shown by W19, namely, of an apparent discrepancy, with the expectation of a monotonic decline of the dust mass content with age, also noted by \citet{2019A&A...626A..11C}. 

An important limitation of the comparative analysis of the cumulative mass distributions is that it does not take into account the properties of the sampled stellar masses. We  attempt to address this limitation in the following sections.

\subsection{\Mdust-\Mstar\ relation}
\label{Sec:MdustMstar}

\begin{figure*}
   \centering
       \includegraphics[width=18cm]{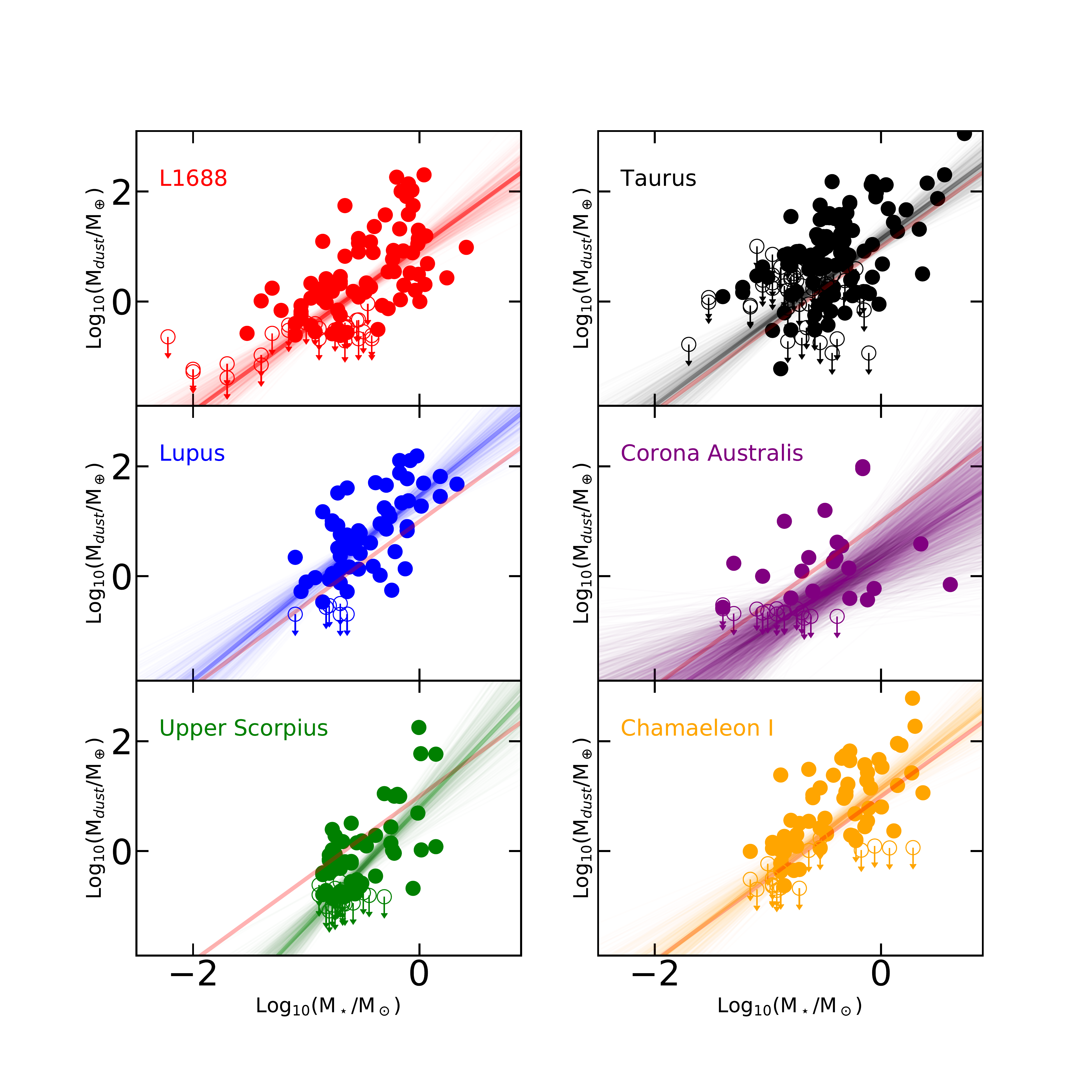}
      \caption{Dust mass \Mdust\ is shown as a function of the stellar mass for the six regions. 
      In each panel the lines show the result of a linear fit performed including the whole range of \Mstar\, as discussed in the text. The orange line shows the best fit for L1688. 
              }
         \label{Fig:Mdust-Mstar}
   \end{figure*}

The distribution of values of the disk mass  versus \Mstar\ is shown in Fig.~\ref{Fig:Mdust-Mstar} for all the regions.
As for L1688,  we performed  for each region,  a 
power-law fit to the data \citep{2007ApJ...665.1489K}  for the whole sample and for \Mstar$>0.15$\Msun (see Appendix~\ref{App:fits}); the results  are shown in Fig.~\ref{Fig:Mdust-Mstar} and Table~\ref{Tab:fit_mdust_mstar}. With the possible exception of Taurus, the inclusion of the lower-mass objects does not change the values of the slopes within the uncertainties.  Slopes vary between $\sim 1$ for CrA and Taurus (we note, however, that the fits are heavily affected by upper limits), $\sim 1.3$ for Chamaeleon and $\sim 2$ for L1688, Lupus and Upper Scorpius. There is no clear trend with the age of the regions, as L1688, Lupus and Upper Scorpius have  similarly steep slopes. We note that the slopes derived for some of the regions are slightly different from those reported by \citet{2016ApJ...831..125P} and \citet{2017AJ....153..240A}, due to a 
combination of the revision of our samples based on the Gaia membership and distance, along with the new photospheric parameters, and the stellar mass range \citep[see also][]{2020ApJ...895..126H}.

\subsection{\Macc-\Mstar\ relation}
\label{Sec:MaccMstar}

\begin{figure}
   \centering
       \includegraphics[width=8.8cm]{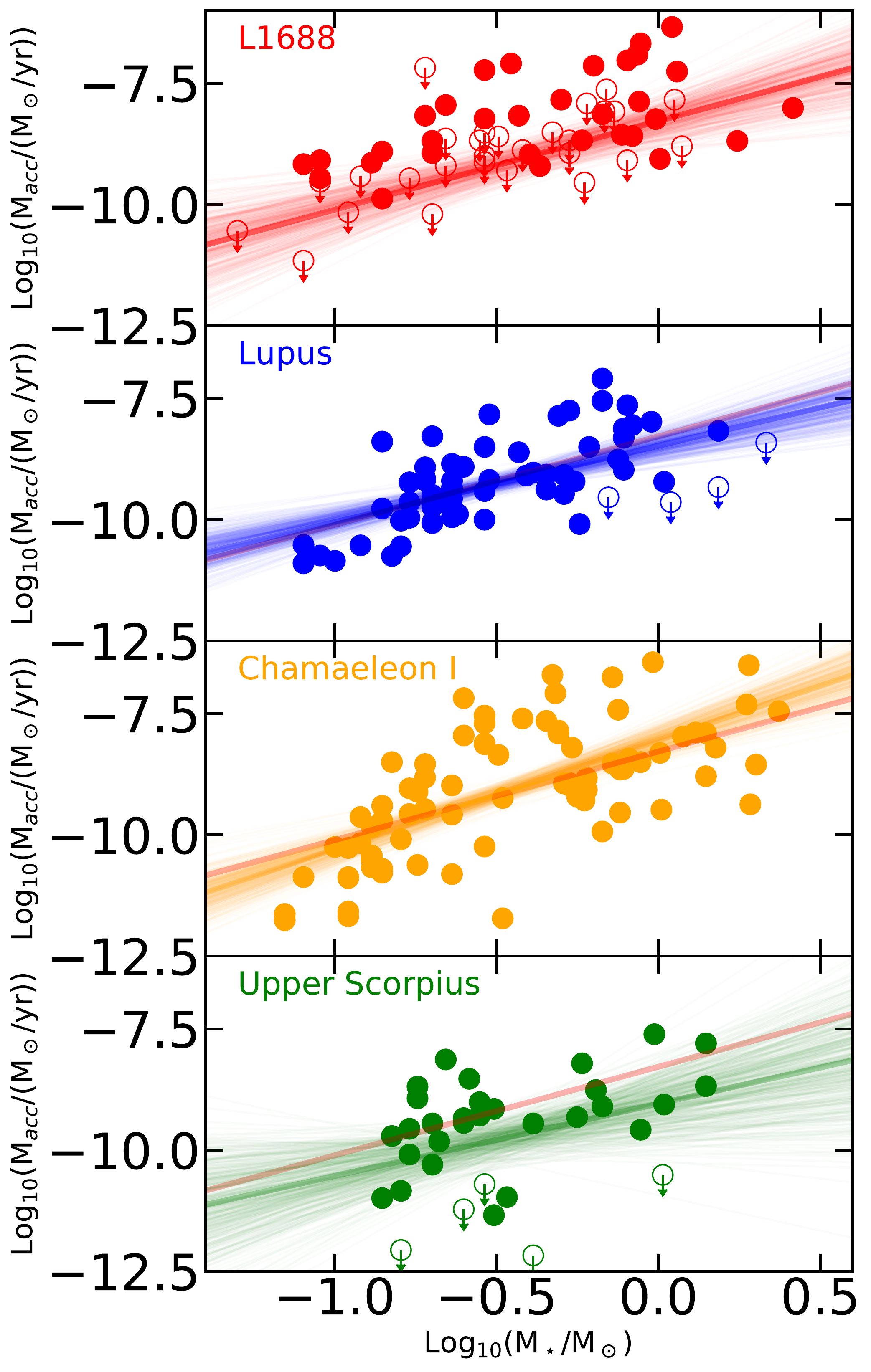}
      \caption{ Mass accretion rate as a function of \Mstar\ for the four  regions for which \Macc\ is known, as labeled.  In each panel we show the result of a linear fit performed including the whole range of \Mstar\, as discussed in the text. The orange line shows the best fit for L1688.
              }
         \label{Fig:Macc-Mstar}
   \end{figure}

Figure~\ref{Fig:Macc-Mstar} plots the mass accretion rate \Macc\ as a function of \Mstar\ for the four regions where homogeneous estimates of \Macc\ are available. Linear fits for the whole samples and the more massive stellar population \Mstar $>0.15$ \Msun\ are shown in Appendix~\ref{App:fits}. The results are given in Table~\ref{Tab:fit_mdust_mstar} and shown in Fig.~\ref{Fig:Macc-Mstar}.
All regions have steep slopes (in the range 1.6--2.3) when the whole sample is considered,  consistent with each other within the uncertainties.  For stellar masses larger than 0.15~\Msun, the nominal values for Lupus and Chamaeleon~I slopes are smaller; however, the difference is not significant considering the uncertainties. \citet{2017AA...600A..20A} and \citet{2017A&A...604A.127M} found that a fit with a break and two slopes is a more accurate description of the data for Lupus and Chamaeleon~I. We do not see evidence that this is the case in the L1688 region \citep[see also][for the case of NGC~1333]{2021arXiv210303863F}. This may point to evolutionary differences in the very-low-mass stars and BDs as compared to stars. In this paper we focus on the evolution of the objects more massive than about 0.15\Msun, as the samples of objects below $\sim$0.1\Msun\ are still strongly affected by incompleteness and small number statistics.

\subsection{\Macc - \Mdisk\ relation and accretion disk lifetimes}
\label{Sec:MaccMdisk}

\begin{figure}
   \centering
       \includegraphics[width=8.8cm]{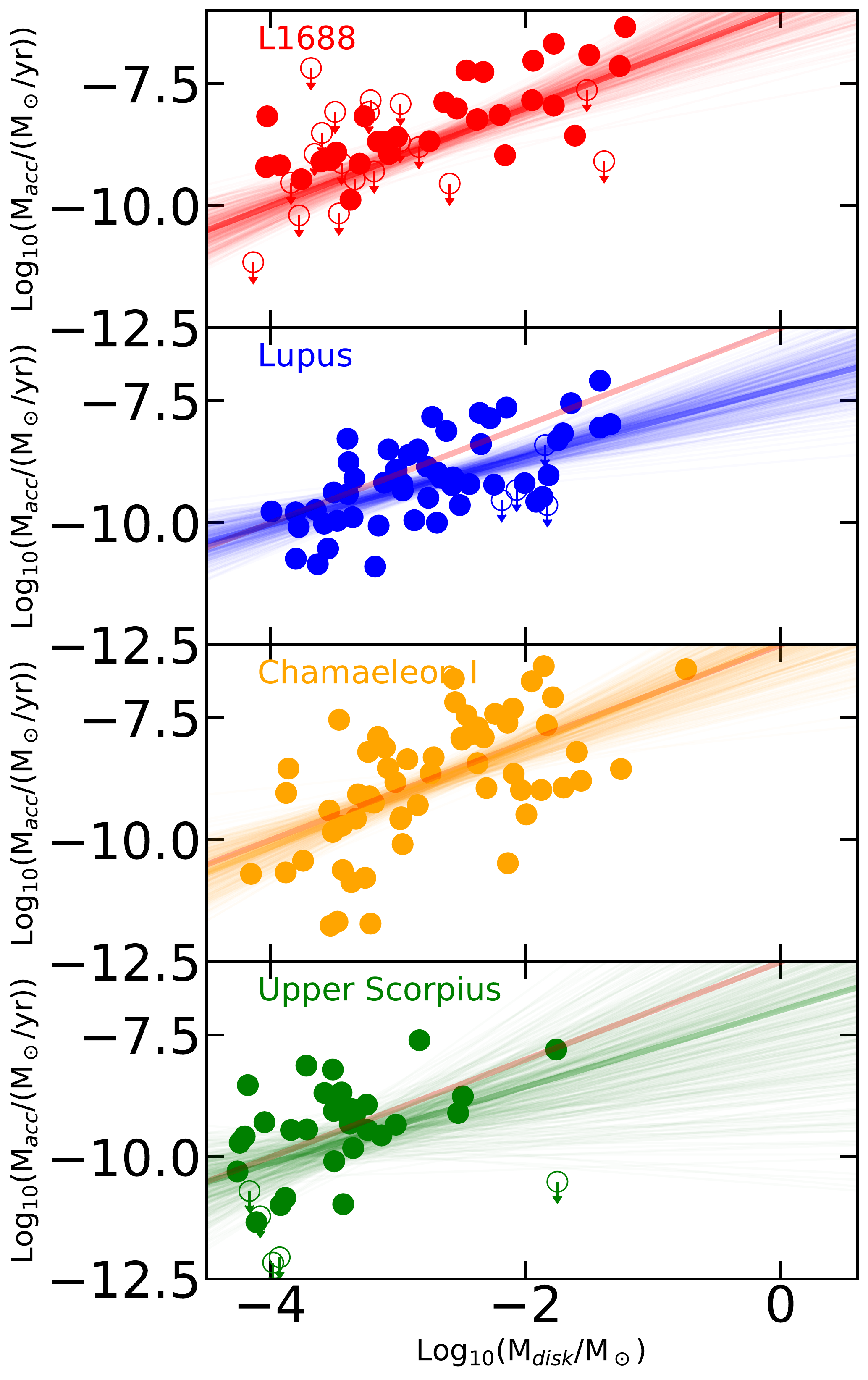}
      \caption{\Macc\  as a function of the disk mass \Mdisk$=100 \times$\Mdust\  for all the objects with measured \Mdisk. In each panel, we show the result of a linear fit performed including the whole range of \Mstar, as discussed in the text. The orange line shows the best fit for L1688.
              }
         \label{Fig:Macc-Mdisk}
   \end{figure}

Figure~\ref{Fig:Macc-Mdisk} shows the relation between \Macc\ and the disk mass \Mdisk\ for all objects with measured \Mdisk\ (upper limits are not included).  
Linear fits are described in  Appendix~\ref{App:fits}. The results show that the trend is about linear for all four regions, with no significant difference between them (see Fig.~\ref{Fig:Macc-Mdisk}), nor for different \Mstar\ intervals. A roughly linear trend was found by \citet{2016AA...591L...3M} in the Lupus star-forming region, by \citet{2017ApJ...847...31M} in Lupus and Chamaeleon~I , and by \citet{2020AA...639A..58M} in Upper Scorpius. We confirm these results and find that this is also the case for L1688.

\begin{figure}
   \centering
       \includegraphics[width=8.8cm]{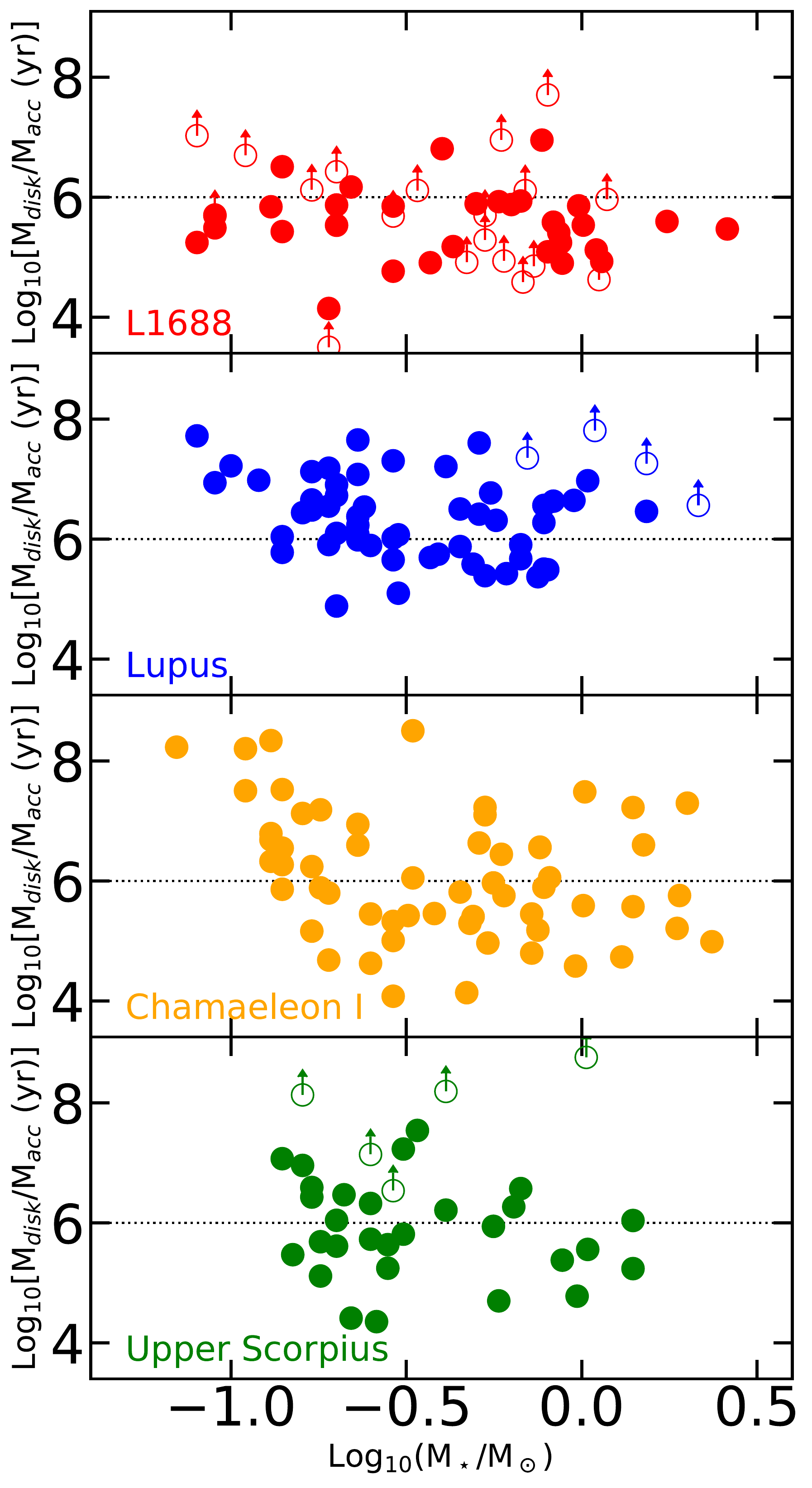}
      \caption{Ratio  \Mdisk/\Macc \ is plotted as a function of \Mstar\ for all objects with measured \Mdisk. The dotted line corresponds to \Mdisk/\Macc=1 Myr.
              }
         \label{Fig:taudisk-Mstar}
   \end{figure}

The ratio \Mdisk/\Macc\ provides an interesting timescale for disk evolution after the first million years, and is plotted in Fig.~\ref{Fig:taudisk-Mstar} as a function of \Mstar. We can see that there is no trend with \Mstar\ for \Mstar > 0.15 \Msun; at lower \Mstar, there is a hint of larger values of this ratio  among Lupus \citep[as noted by][]{2020A&A...633A.114S} and Chamaeleon I objects, but not in L1688. However, there are very few objects at these low masses, and a large number of \Macc\ upper limits in L1688, which, taken all together, limit the significance of this result. 

\begin{figure}
   \centering
       \includegraphics[width=6.cm]{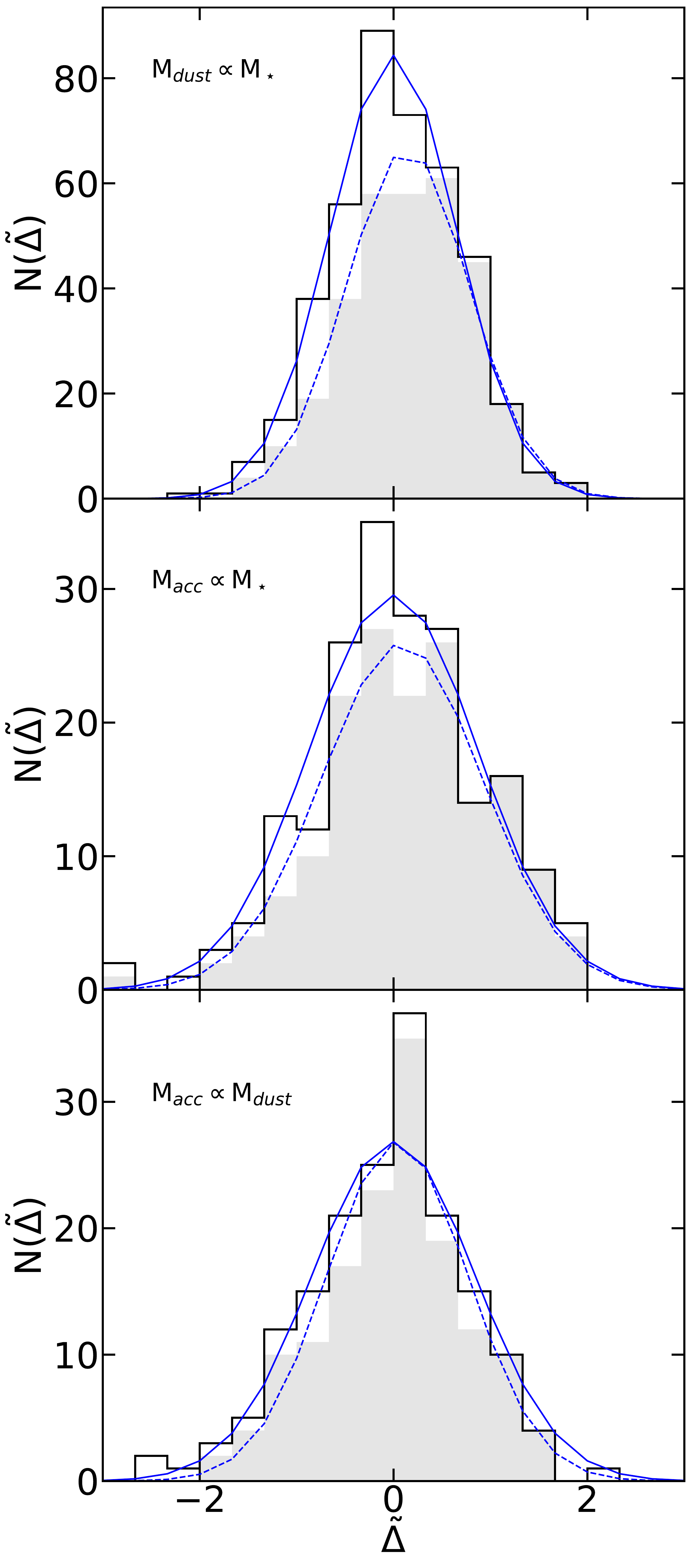}
      \caption{Distribution of the dispersion from the individual best-fitting relations for  of all regions combined (see text). The dispersion (in dex units, $\tilde{\Delta}$) is computed as the distance of the individual values of \Mdust\ and \Macc\ from the best-fit relation with \Mstar\ (top and middle panels) or \Mdisk\ (bottom panel), we include all objects with \Mstar $>$0.15 \Msun. The solid black lines show the results when upper limits are treated as actual measurements, gray shaded areas show the results when they are excluded. The blue solid and dotted lines show the Gaussian fits to the two distributions. 
              }
         \label{Fig:FitDispersions}
   \end{figure}

\subsection{Dispersion around the linear fits}  
\label{Sec:FitDispersions}

All regions show a broad range  of \Mdust\ and \Macc\ values  for the same \Mstar. Inspection of Figs.~\ref {Fig:Mdust-Mstar}, \ref{Fig:Macc-Mstar}, and \ref{Fig:Macc-Mdisk} shows  that in all cases the observed values are distributed over an interval of at least two dex, with no obvious trends.  We quantify this dispersion by measuring the 
distance (in dex units, $\tilde{\Delta}$) of the values from the best fits at fixed values of \Mstar. We consider the sample of \Mstar$\ge 0.15$\Msun\ and add together the results of all the regions. In this way, we correct for potential systematic differences among regions and significantly improve the statistical significance of the results, which are shown in Fig.~\ref{Fig:FitDispersions} for \Mdust\ (top panel) and \Macc\ (middle panel), respectively.

Both distributions are   well described by a Gaussian   with a width of  $\sigma =$ 0.65 and 0.87 for \Mdust\ and \Macc, respectively, when upper limits are included and treated as measurements.  Figure~\ref{Fig:FitDispersions} shows also the results when they are excluded (dotted lines); the distributions are still well described by Gaussians with similar widths ($\sigma$ =0.64 and 0.84, respectively).  Upper limits are concentrated among objects below the fitted relations (negative values of $\tilde{\Delta}$), but do not change the overall distribution too much. 

Finally, we have performed a similar analysis for the dispersion of the observed values of \Macc\ for the same \Mdisk\ (Fig.~\ref{Fig:FitDispersions}, bottom panel). The distribution is again well represented by a Gaussian, with $\sigma$= 0.84 and 0.73 when upper limits are included or not,  similarly to that of the other two distributions. 

\section {Discussion}
\label{sec:disc}   

In this section, we discuss the disk properties defined in the previous sections as a function of the characteristic age of each region.
We note that this approach assumes that one region "ages" into another one and that other differences, such as the initial conditions under which disks form or the environment in which they evolve are irrelevant.


\subsection {Time evolution of dust mass and mass accretion rates}
\label{sec:time_evolution}

\begin{figure}
   \centering
       \includegraphics[width=8.8cm]{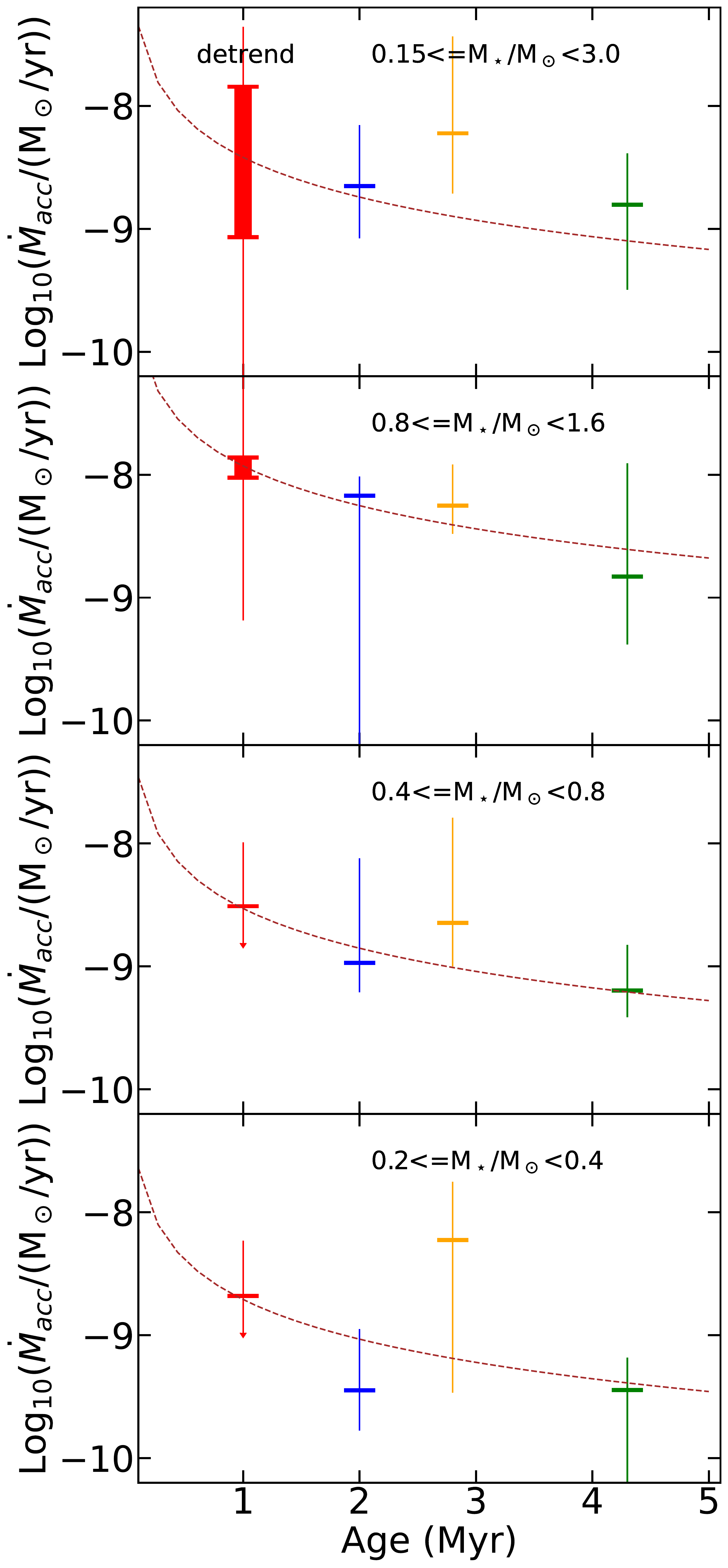}
      \caption{\Macc\ medians and percentiles as a function of age for L1688 (red), Lupus (blue), Chamaeleon~I (yellow), and Upper Scorpius (green). From top to bottom: for the full range of stellar masses $0.15\le M_\star/M_\odot<3.0$ (after removing the trends with $M_\star$ as derived in Appendix~\ref{App:fits}), and for three mass bins, as labeled. In each case, the dotted line shows the trend \Macc$\propto t^{-1}$, normalized to the median value for L1688. The filled box shows the possible range of the median value, following the treatment of upper limits described in the text.
              }
         \label{Fig:Macc_age}
   \end{figure}

In order to control how the results may be affected by the strong dependence of both \Mdisk\ and \Macc\ on \Mstar,  we both use the full 0.15--3 \Mstar\ interval (removing the dependencies on $M_\star$ as derived in Appendix~\ref{App:fits}), along with narrow intervals of \Mstar,\ and we compute the median and quartile values for the various disk (sub-)samples. The results are shown as a function of the age of the region in Figs.~\ref{Fig:Macc_age} and \ref{Fig:Mdust_age}.




In all cases, we compute how medians and percentiles are affected by upper limits by comparing the two extreme cases:  assuming that the true value of the quantity is equal to the estimated upper limit, or that the true value is much smaller than the smallest measured value in the full sample. The two values of the medians are often very similar; when they are not, the range between the two extremes appears as a filled box in the figure. In some case, the number of upper limits is larger than the number of detections in the given \Mstar\ interval. 
In such case the lower extreme of the median range cannot be defined by our procedure and we show the median as an arrow. The 75\%\ percentile is always  defined.  
 
A close inspection of the results shows that it is possible to identify an overall temporal trend, similar for both \Mdust\ and \Macc: a slow, linear decrease with time ($\propto t^{-1}$), as shown in Fig.~\ref{Fig:Macc_age}. 
This temporal evolution is expected for viscous disks (under the assumption that \Mdust\ traces the total disk mass). \citet{2016ARA&A..54..135H} report an almost linear time decrease of \Macc$\propto t^{-1.07}$ in a large sample of individual stars from various star forming regions  with masses in the interval 0.3--1.0 \Msun and  ages, derived from the location of the individual stars on the HR diagram, ranging from $\sim 0.1$ to $\sim 20$ Myr. 

As  shown in Fig.~\ref{Fig:taudisk-Mstar}
there is no evidence for a time evolution of the ratio $\tau= $\Mdisk/\Macc. If we assume a simple viscous evolution of the disk populations, we expect that when the age of the disk population is much larger than the viscous timescale, $\tau$ will increase with the age of the region \citep{2017MNRAS.472.4700L}. This is, in fact, not the case and $\tau$ remains constant at $\sim$1~Myr over an interval of ages of several Myr, presenting a challenge for  simple viscous evolution models.

\subsection{Disk mass and dust mass}
\label{Sec:diskmass}

The similarity of the time evolution of  \Mdust\ to that of \Macc\ indicates that \Mdust\ is  a good overall, first-order tracer of \Mdisk, as already noted by \citet{2016AA...591L...3M}.
The disk mass  is computed assuming a constant, canonical factor of \Mdisk = 100$\times$ \Mdust. This is just a normalization factor. In fact, the overall trend, which is similar to that of \Macc, does not change as long as  \Mdisk/\Mdust\ remains roughly constant over the time interval spanned by the regions studied in this paper. 
If planet formation occurs at an earlier stage and significantly depletes the disk of its grain population, then our assumption of \Mdisk/\Mdust=100 in the 1~Myr old Class II population of L1688, Taurus and Corona Australis  may  grossly underestimate  the total disk mass at that age. In this case,  the true timescale (\Mdisk/\Macc) will be proportionally higher than the $\tau \sim 10^6$~yr indicated in Fig.~\ref{Fig:taudisk-Mstar}. However, the constancy of $\tau$ in the following evolution remains. The results of this paper do not give any direct constraint on the value of \Mdisk/\Mdust\ in the youngest Class II samples, but suggest that, whatever the initial value, the gas and dust evolution is not massively decoupled in the following few million years, at least in disks that survive the passing of time.

\subsection {Dispersions}
\label{Sect:dispersion}

The three relations we  analyzed are dominated by the large dispersion of points  around the best-fitting relations (see Fig.~\ref{Fig:FitDispersions}), well above the observational uncertainties \citep[see e.g. discussion in][]{2020AA...639A..58M,2017ApJ...847...31M}.  The distribution of the distances from the best fits is smooth and well described by Gaussian functions. The dispersion is a characteristic of the whole populations and it is not influenced by a limited number of peculiar objects; therefore, it is therefore related to a dispersion of the underlying physical  processes. The dispersion in the \Macc --\Mdisk\ relation is similar to the one in the other two relations, namely \Mdisk\ versus \Mstar\ and \Macc\ versus \Mstar.  A possible interpretation is that stars of the same age and mass can accrete  at very different mass accretion rates, even when their disk mass is the same \citep[e.g.][]{2006ApJ...645L..69D,2017ApJ...847...31M,2020AA...639A..58M}. 
The origin of this is still not understood, and there have been a number of different suggestions (initial conditions, time variability, magnetic fields, disk dissipation processes), none of them fully satisfactory. 
Alternatively, it has been suggested that disks of the same gas mass can have very different dust evolution causing an additional dispersion in the relation between the (dust-derived) disk masses and mass accretion rates \citep{2020MNRAS.498.2845S}. Our results (Fig.~\ref{Fig:FitDispersions}) may provide quantitative constraints to both interpretations.

\subsection{\Mdust\ excess in Lupus and Chamaeleon~I}
\label{Sec:md_lup_cha}

\begin{figure}
   \centering
       \includegraphics[width=8.8cm]{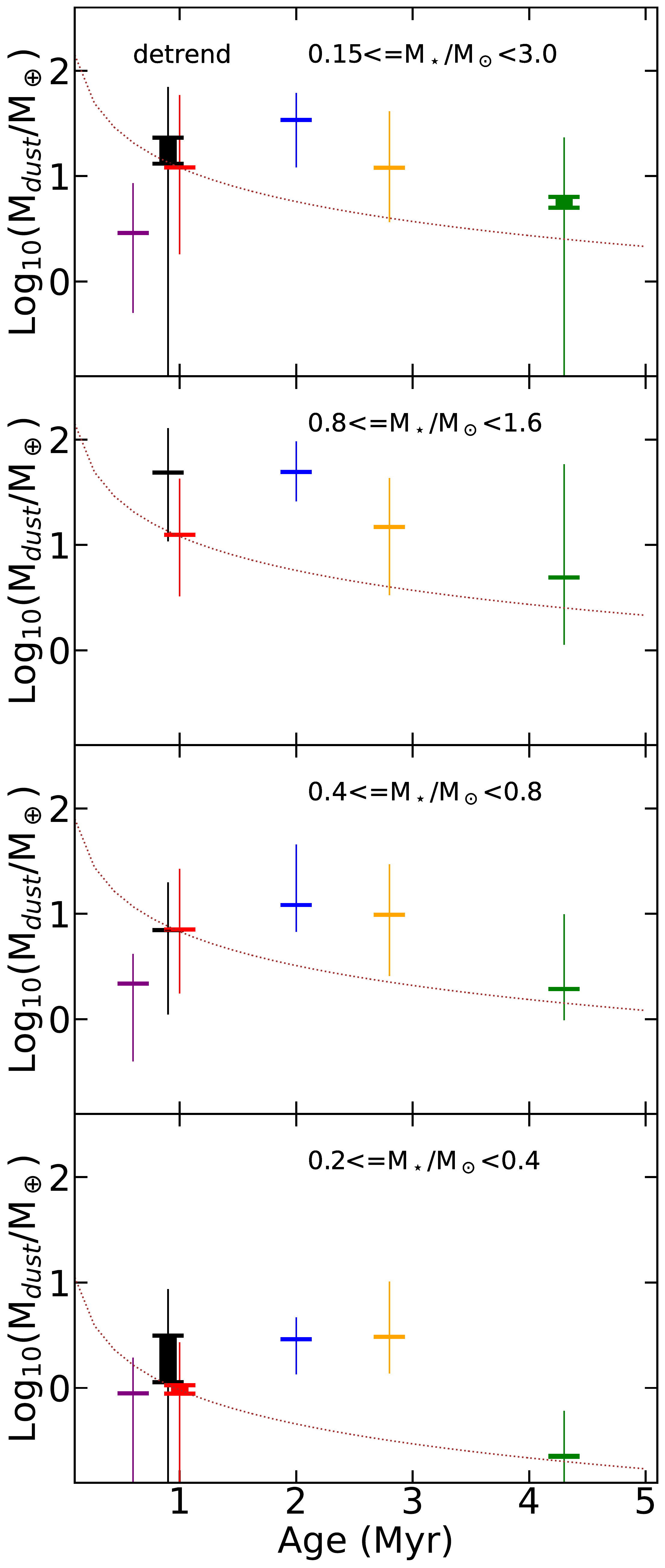}
      \caption{\Mdust\ medians and percentiles as a function of age for L1688 (red), Corona Australis (purple), Taurus (black), Lupus (blue), Chamaeleon~I (yellow), and Upper Scorpius (green). From top to bottom: For the full range of stellar masses $0.15\le M_\star/M_\odot<3.0$ (after removing the trends with $M_\star$ as derived in Appendix~\ref{App:fits}), and for three mass bins, as labeled. In each case, the dotted line shows the trend \Mdust$\sim t^{-1}$, normalised to the median value for L1688. The filled box shows the possible range of the median value, following the treatment of upper limits described in the text.
              }
         \label{Fig:Mdust_age}
   \end{figure}

Figure~\ref{Fig:Mdust_age}  shows an interesting effect, on top of the time dependence defined by the dotted line, namely, displaying how the two regions with ages of $\sim 2-3$Myr (Lupus and Chamaeleon~I) have, on average, more massive disks than expected for their age. The difference is a factor of $\sim 3$ and it is rather uncertain. We note that it appears to be more pronounced for low-mass stars. It is, however, potentially very interesting, as dynamical models of the evolution of dust, planetesimals, and planetary populations in disks predict that the total mass of grains  detectable at mm wavelengths may not decrease monotonically with time, but increase after the formation of sufficiently large planets, as these excite a  population of planetesimals into eccentric orbits and trigger collisional fragmentation 
\citep{Turrini2012,2019ApJ...877...50T,2019A&A...629A.116G}.
If this is indeed the case, then our results suggest that the peak of this process should occur at  2-3Myr, and point to an earlier phase of planet formation. In this view, planets form on timescale of about 0.5-1Myr, or even less. Objects such as L1688, Taurus, and Corona Australis may be at a stage where planets have already formed, but the creation of a significant population of secondary grains is just beginning. Their dust content is still dominated by the remaining fraction of the original solid population. 
Over the following 1-2~Myr, collisional fragmentation  of planetesimals increases the dust mass; further in time, this dust production stops and the disk dust mass declines.

This view is consistent with the current idea that planet formation has to occur at very early ages, before the age of the youngest populations discussed in this paper. We note that this generation of a secondary population of dust at 2-3 Myr is not large enough to change the overall trend of the time evolution, as discussed in Sec.~\ref{sec:time_evolution}.

\subsection{Disk dispersal processes}
\label{Sec:diskpop}

If, as discussed in Sec.~\ref{sec:time_evolution}, \Macc\ and \Mdisk\ decrease with time approximately as $t^{-1}$,  after 5 Myr the disk mass and mass accretion rate would only be reduced by a factor of 5. This would result in very long dispersal times, much longer than the typically observed value of few Myr \citep{2008ApJ...686.1195H}. The fact that Class~III objects show mm emission similar to that of field debris disks \citep{2021MNRAS.500.4878L,2021arXiv210405894M} suggests that disks may need to dissipate very fast, much faster than the ages of those we observe in the surviving Class~II populations. The most likely candidates for these fast dissipation processes are photoevaporation or planet formation. Population synthesis models predict a clear tail of low values in the \Macc\ vs. \Mdisk\ relation, as most of the disk dispersal and disk planet interaction processes predict a faster decline of \Macc\ compared to \Mdisk. This is not observed in the present study (see Fig.~\ref{Fig:FitDispersions}) and 
it is possible that the short timescale of these processes makes it very difficult to find evidence for this in the properties of the surviving disks \citep[see also][]{2019A&A...631L...2M}.
Our large samples, which span a large range of ages, offer, in principle, the possibility of identifying the tail of objects in the dispersal phase, but we do not find any clear evidence for such a tail (see Sect.~\ref{Sec:FitDispersions}). In Fig.~\ref{Fig:FitDispersions} (bottom panel), the few objects at $\tilde\Delta <-2$ could be further investigated as potentially residing in the phase of dispersal.

We conclude that the analysis of bulk population properties of  Class~II disks  does not tell us how disks disappear; these follow the slow evolution that is consistent with viscosity or wind-driven accretion, not the catastrophic events that lead to the disk dissipation on short timescales. The potential impact of different planetary architectures and dust evolution on the timescales of disk dissipation is currently being discussed and may be a dominant factor in explaining the observed dispersion of properties \citep[see Sect.~\ref{Sect:dispersion}, and ][]{2020MNRAS.498.2845S,2020A&A...635A.105P,2021AJ....162...28V,2021arXiv210405894M}.

\section{Summary and conclusions}
\label{concl}

This paper provides a description of disk masses and mass accretion rates in the $\rho$-Oph region L1688, one of the richer and younger star-forming regions in the Solar neighbourhood. Our selection criteria require that the objects have been observed with ALMA (W19),  have distance and membership properties taken from Gaia (EL20), and an estimate of spectral type and extinction. We estimated \Teff, \Lstar, and \Mstar\ for each object and derived a characteristic age of the population of $\sim 1$~Myr. The relations between \Mdisk\ and \Mstar, \Macc\ and \Mstar, and \Macc\ and \Mdisk\ are qualitatively similar to those observed in other regions,  with a roughly linear trend with slopes $\sim$ 1.8--1.9 for the first two relations and $\sim 1$ for the third (for \Mstar $>0.15$ \Msun). All three relations have large dispersion around the best fit. 

The main interest in studying L1688 is to put it in the context of the other regions studied so far, and to characterize, in particular, the disk evolution over time. To this end, we  re-analyzed the data for Corona Australis, Taurus, Lupus, Chamaeleon I and Upper Scorpius, using the same selection criteria and analysis as for L1688. The stellar and disk properties are given in the Appendix~\ref{App:tabs}. 
The results confirm the results of previous analyses in the literature, with  differences due to different selection criteria and assumptions, that do not affect the global picture.

We note that our selection of Class II/Flat disks limits our results to the evolution  from the time the star-disk systems emerge from their parental core to the time they lose the near or mid-IR excess used in disk classification. The star-forming regions in this paper range in age from  about~0.5 to~5~Myr.  When ordered according to the characteristic age of each region, there is evidence of a relatively slow decline of \Macc, roughly as $t^{-1}$. A similar trend is observed when plotting the properties of a sample of  individual stars as function of the stellar age \citep[]{2016ARA&A..54..135H} and is expected if disk evolution is controlled by some kind of viscosity.

The behavior of \Mdust\ over time is more complex. Assuming the same conversion factor from millimeter flux density to dust mass, 
the values of \Mdust\ in L1688 (age $\sim 1$Myr) and in Upper Scorpius (age $\sim 4.5$Myr) show a decline that is consistent with the same trend as observed in \Macc.
However, an interesting aspect of the \Mdust\ behavior is that it is not monotonic with age, but shows an increase of a factor of $\sim$3 at the age of 2-3~Myr above the overall time trend. A similar behavior is predicted by   some theoretical models  where the earlier ($<$1Myr) formation of planets excites a population of planetesimals into eccentric orbits,  subsequently triggering a collisional fragmentation cascade \citep{Turrini2012,2019ApJ...877...50T,2019A&A...629A.116G}. 
This result is very interesting as it is a potential confirmation of early planet formation. Nonetheless, we note that our observations suggest that the amount of millimeter detected dust  is only a factor of $\sim$3 larger than the value expected from the evolutionary trend. In turn, this means that the 
total disk mass estimates, based on a conversion from the millimeter continuum flux, are probably qualitatively correct within the same factor.

All the relations have large dispersions. To minimize the limitations induced by the relatively small samples, for all the regions we combined the dispersions from each of the individual fits.
We find that the dispersions of measurements around each of the three relationships \Mdust\ versus \Mstar, \Macc\ versus \Mstar, and \Macc\ versus \Mdisk\
have continuous distributions with   a roughly log-normal shape and similar widths ($\sim \pm 0.8$ dex), with no outliers.
In particular, in the \Macc\ versus  \Mdisk\ relation, there is no evidence of a tail of objects with low \Macc. This confirms  the need for a  rapid disk-dispersal mechanism, such as photoevaporation or planet formation, which, however, do not seem to significantly affect  the disk properties considered in this paper {\it during the  Class II/F phase}.

When compared to the predictions of viscous disk evolution, we confirm an almost linear decrease of both \Macc\ and \Mdisk\ over time. However, viscous models encounter  a number of difficulties, already noticed in the literature \citep[eg][]{2017ApJ...847...31M,2020AA...639A..58M}, which our study confirms. Among them, there is the large dispersion in the 
\Macc --\Mdisk\ relation, which does not change over time, along with the fact that the ratio  of these two quantities is 
 constant in time.

The origin of two properties remains puzzling: the steep dependence of \Macc\ and \Mdisk\ on \Mstar\ and  the cause of the large dispersion in the three relations analyzed in this paper; in particular that of the \Macc\ versus \Mdisk\ relation. These dispersions could be related to a particular combination of different effects, including initial conditions, dust evolution, and disk dispersal properties \citep[eg][]{2006ApJ...645L..69D,2017MNRAS.472.4700L,2020MNRAS.498.2845S,2020MNRAS.492.1120S}.

An important caveat is that even the relatively large and well-characterized samples discussed in this paper are still limited in a number of ways. Two major limitations are (still) the sensitivity limits and the completeness of the samples. For example, values of \Macc\ in L1688 come from a relatively old work \citep{2006AA...452..245N}, while about 50\%\ of the measurements are upper limits and many of the recently classified members by EL20 were not included. Even the extensive ALMA survey has a large number of upper limits and is incomplete. The same is true for Upper Scorpius.

To attain progress in the understanding of  disk populations evolution and dissipation, it is necessary to overcome the limitations outlined above. It is equally important  to dedicate significant amounts of ALMA time to obtain systematic measurements of molecular gas properties, as well as disk radii. Extending similar studies to the earlier  (e.g., Class 0 and I) and later (Class III) phases of disk evolution is necessary to explore the  planet formation phase, as well as disk dissipation.

\begin{acknowledgements}
This work was partly supported by the Italian Ministero dell Istruzione, Universit\`a e Ricerca through the grant Progetti Premiali 2012 – iALMA (CUP C$52$I$13000140001$), 
by the Deutsche Forschungs-gemeinschaft (DFG, German Research Foundation) - Ref no. 325594231 FOR $2634$/$1$ TE $1024$/$1$-$1$, 
and by the DFG cluster of excellence Origins (www.origins-cluster.de). 
This project has received funding from the European Union's Horizon 2020 research and innovation programme under the Marie Sklodowska-Curie grant agreement No 823823 (DUSTBUSTERS) and from the European Research Council (ERC) via the ERC Synergy Grant {\em ECOGAL} (grant 855130), and the ERC Advanced Grant {\em The Dawn of Organic Chemistry} (grant 741002).
K.M. acknowledges funding by the Science and Technology Foundation
of Portugal (FCT), grants No. IF/00194/2015, PTDC/FIS-AST/28731/2017,
 and UIDB/00099/2020. IdG-M is partially supported by MCIU-AEI (Spain) grant AYA2017-84390-C2-R (co-funded by FEDER).
\end{acknowledgements}


\bibliographystyle{aa} 
\bibliography{OfiuconeII.bib} 


%

\begin{appendix} 
\section {ALMA 0.89mm observations of BDs and very low mass objects in L1688}

\label{sec:Alma-BDs}

\citet[][T16]{2016AA...593A.111T} observed 17 BDs and very-low-mass objects in L1688 at 0.89mm. 
We note  that 1 of the 18 published in T16 (ISO-Oph 164) was, in fact, an erroneous observation (while the source coordinates in Table~1 in T16 are correct, the source really observed with ALMA was GY92-317, located at 16:27:40.10 $-$24:38:36.46). 
Nine additional sources were observed as part of the ALMA project {\tt ADS/JAO.ALMA~2017.1.01243.S} on April 1$^{st}$, August 17$^{th}$, and 18$^{th}$, 2018. A total of 44 Antenna Elements were available during the observing sessions. We used the ALMA Band 7 receivers tuned to a frequency of $\sim$338~GHz, the total effective bandwidth usable for continuum was approximately 6~GHz and the 13CO(3--2) line was covered in one of the spectral windows of the correlator. Standard calibration was performed by the ESO ALMA Regional Centre, and the flux density scale is expected to be accurate within 5\%. The total time on source was about 12min per target. Peak, integrated fluxes, and upper limits are computed as described in T16.

The 0.89mm fluxes and upper limits  are  displayed in Table~\ref{tab:Alma-BDs} for the total sample of 26 objects; data from T16 are marked as such in the last column of the table. All objects but ISO-Oph42 are classified as member or candidate member by EL20; distances, spectral types, and extinction come from their paper. For ISO-Oph42, stellar parameters are from \citet{2012ApJ...744..134M}, as given in T16, corrected for the new Gaia distance. Stellar luminosity and mass are computed as described in Sect.~\ref{oph_samp}.

\begin{table*}
\tiny
\begin{center}
\caption[]{VLMO with ALMA 0.89mm data}
\label{tab:Alma-BDs}
\begin{tabular}{l l l l l l l   l l l l l  }
\hline

Name  &            $\alpha$ &           $\delta$&             d&   SpT&      T$_{\rm eff}$ &   A$_{K}$ & Log$_{10}(L_\star)$ &   M$_\star$ &  F$_{peak}$ & F$_{tot}$ &   Ref  \\
           &            (J2000)&         (J2000)&       (pc)&    &    (K)&   (mag)& (L$_\odot$) &  (M$_\odot$)&  (mJy/beam)&   (mJy)&     \\
\hline
 CFHTWIR-Oph-16 &     16:26:18.5804 &  -24.29.51.8544 &    139.4&       M8&      2710 &  1.51  & -2.0 &    0.03&           1.0$\pm$    0.08&     1.0 $\pm$ 0.1&    T16 \\
ISO-Oph23   &       16:26:18.8158&  -24.26.10.5216 &   139.4 &   M6.5&    2930&   1.02&   -1.5  &    0.08&     1.30 $\pm$  0.16 &    1.50$\pm$  0.23&    T16 \\
ISO-Oph26  &    16:26:18.9823&  -24:24:14.2596  &  139.4    & M6  &   2990 & 2.86&         -0.68&    0.18   &  46  $\pm$   0.06 &    56.4 $\pm$   0.2 &        \\
ISO-Oph30  &        16:26:21.5319 & -24:26:01.3934 &  136.6 &   M5.5&    3060&  0.46&   -1.16&  0.14 &      4.20$\pm$   0.15&     4.8$\pm$   0.17& T16 \\
ISO-Oph32  &        16:26:21.9016&  -24:44:40.0929 &   151.3 &      M7.25&  2840&  0.27 &  -1.18& 0.08  &        1.620$\pm$  0.099  &  1.80$\pm$  0.11    &  T16 \\
ISO-Oph33$^{1}$    &   16:26:22.2644  & -24:24:07.556  &   139.4   &    M7  &   2880 &  0.88  & -2.23 &  0.06 &  1.3 $\pm$   0.18  &   1.30 $\pm$ 0.15    &   T16\\
ISO-Oph35  &        16:26:22.9557 & -24:28:46.15 &     139.4     &    M6  &   2990 & 1.48&  -1.08& 0.12&           -- &     $<$ 0.9 &     T16 \\
CRBR-2322.3-1143 &  16:26:23.8142& -24:18:29.0016&    139.4    &   M5.5&   3060 &  0.99 &  -2.06&  0.10 &         --      &      $<$ 0.6 &       T16\\
ISO-Oph42  &       16:26:27.8166& -24.26.41.8665  &  165.86   &   M5&     3125 &  0.02 &   -2.12 &0.13  &   3.61 $\pm$ 0.19  &   4.2  $\pm$ 0.23 & T16 \\
CFHTWIR-Oph-31 &    16:26:37.8135& -24:39:03.1352 &  139.4   &   M5.5&  3060 &   1.96& -1.78&0.11&           1.6 $\pm$   0.05&     1.41$\pm$   0.05&     \\
GY92 141  &        16:26:51.2908& -24:32:42.0177 &  142.7 &   M8.5 &  2550 &  0.15  &  -2.54 &0.02  & 0.13$\pm$  0.05  &   $<$ 0.28   &     \\
CFHTWIR-Oph-58   &  16:27:05.9268& -24:18:40.2156  &  139.4 &   M8 &  2710 &0.8 &   -2.70& 0.04   &  0.11 $\pm$ 0.05   &  $<$ 0.26     & \\
GY92 202  &        16:27:05.9762& -24:28:36.3252&    139.4 &    M4.5&   3200 &  1.86&  -1.50& 0.18&         --     &   $<$  0.5 &    T16\\
ISO-Oph102  &       16:27:06.5925& -24:41:48.883 &   142.1 &    M5.5   & 3060& 0.44& -1.09& 0.14&     3.66 $\pm$ 0.19   &   3.8 $\pm$  0.19 &   T16\\
CFHTWIR-Oph-66$^{3}$  &   16:27:14.3404& -24:31:31.9260  &  139.4   &    M7.75 & 2750& 1.37&        -2.65& 0.05&       6.1  $\pm$ 0.06   &   6.7 $\pm$  0.07 & \\
CFHTWIR-Oph-77   &  16:27:25.6440& -24:37:28.5168   & 139.4  &    M9.75&  2260&  0.94 & -3.02  &0.02&      -0.16 $\pm$ 0.05& $<$0.15 &\\
ISO-Oph138 &      16:27:26.2233& -24:19:22.9980&    139.4  &   M6&  2990& 1.6& -1.60& 0.09&     1.95$\pm$  0.085&     2.0$\pm$   0.086&     T16\\
GY92 264   &     16:27:26.5817& -24:25:54.3936  &  139.4  &    M8    & 2710 &  0.04 &  -1.74    &0.04&     3.965$\pm$ 0.090   &  4.1 $\pm$  0.093&  T16    \\
CFHTWIR-Oph-90  &   16:27:36.6115& -24:51:36.0900  &  139.4    &   L0  &   2250 & 0.25 &  -2.65  &0.01   &    0.07$\pm$  0.05  & $<$  0.22 &     T16\\
ISO-Oph160   &      16:27:37.4272 &-24:17:54.955  &   139.4 &       M5.5&   3060&   0.78& -1.46& 0.12 &   6.13 $\pm$ 0.20 &     7.60 $\pm$0.24 &  T16 \\
GY92 320      &    16:27:40.8482  &-24:29:00.7868  &  139.9  &   M7.5 & 2753&   0.27 &  -2.15 & 0.04&     -- &     $<$ 0.4   &T16\\
CFHTWIR-Oph-98   &  16:27:44.226  & -23:58:52.14   &   139.4  &   M9.75 & 2260 &  0.48&  -2.96 &0.01 &      0.06$\pm$  0.05 &    $<$ 0.21&     \\
ISO-Oph176 &     16:27:46.2918&  -24:31:41.2646&    138.9 &    M6&   2990&  1.01& -1.11& 0.12&      --  &   $<$  0.5&         T16 \\
ISO-Oph193  &       16:28:12.723&   -24:11:35.7237 &   151.7  &  M4.5 &   3200&  0.93& -1.00& 0.20&     7.82$\pm$  0.20  &    8.7$\pm$   0.22&    T16   \\
CFHTWIR-Oph-107 &   16:28:48.7135& -24:26:31.7096 &  159.6  &   M6.25&  2960& 0.22 & -2.02& 0.07&         0.13 $\pm$ 0.05  &   $<$ 0.28  &      \\
\hline
\hline
\end{tabular}
\end{center}
\end{table*}

\begin{figure}
   \centering
       \includegraphics[width=9cm]{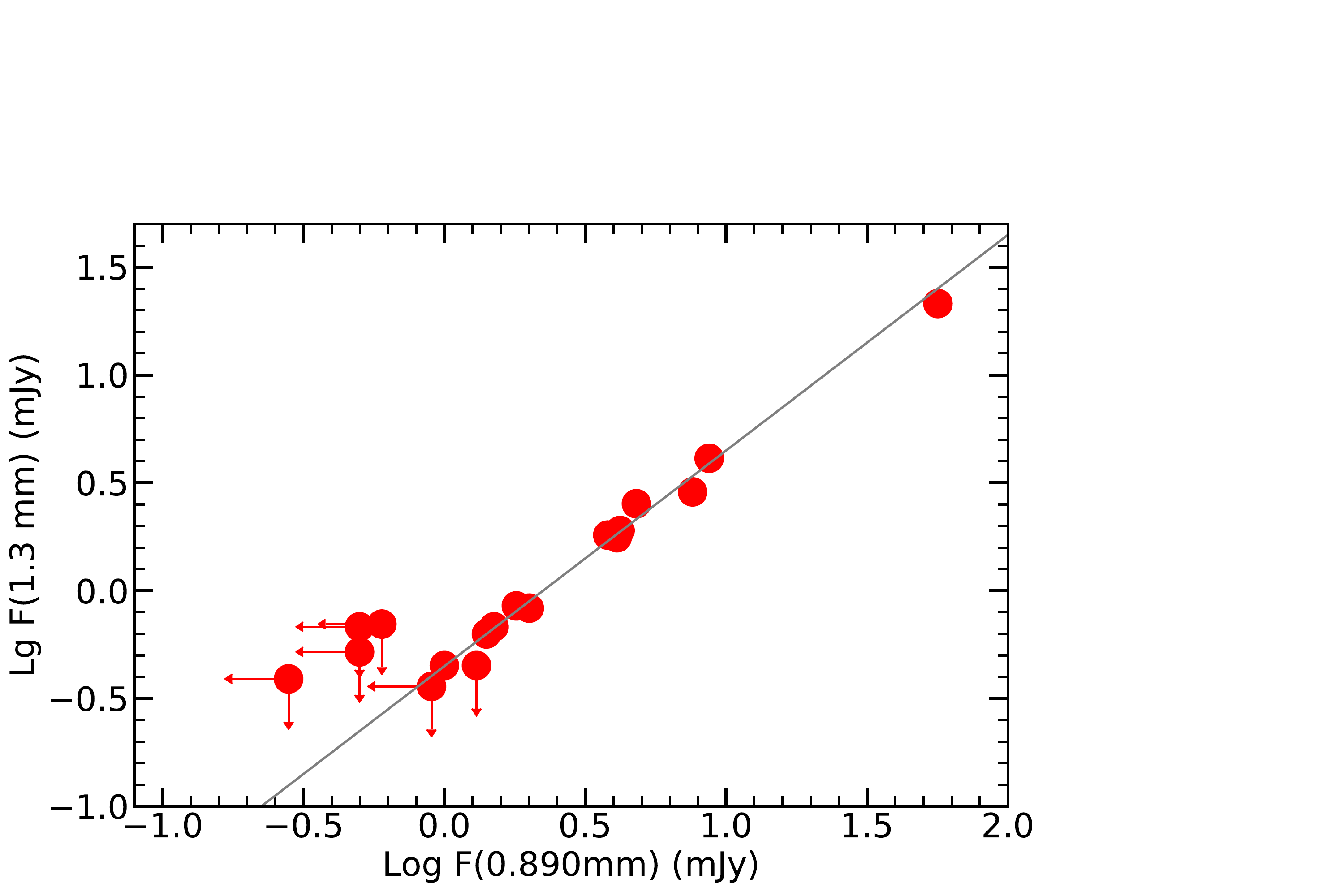}
      \caption{Observed 1.3mm versus 0.890mm flux for the L1688 objects with measurements at 1.3mm  and  0.98 mm.
              }
         \label{Fig:spectralindex}
   \end{figure}

W19 gives measurements or upper limits for 19 of the objects in Table~\ref{tab:Alma-BDs}.
Figure~\ref{Fig:spectralindex} plots the 1.3mm flux versus the 0.890 flux for all the objects with both measurements; it shows that there is a very tight relation, with a ratio $F_{1.3mm}/F_{0.89mm}$=0.44.  The corresponding spectral index is $\alpha_{mm}=2.2$.

We caution that, in order to make a comparison with the recent literature, in this paper we adopt a conversion factor between observed millimeter flux and dust mass using a constant dust temperature of 20~K. This is different from the assumption of T16, who used the prescription T$_{dust}\sim L_\star^{0.25}$ \citep[following][]{2013ApJ...771..129A}. The derived \Mdust\ values are systematically lower than those in T16, by a factor of 3--10; see also the discussion on the adopted value of $T_{dust}$ in T16 Appendix~A.  However, these different assumptions do not affect the slope of the \Mdust -\Mstar\ relation. 


\section{$\rho$--Ophiuchi/L1688 data}
\label{App:tabs}

Parameters of the Class~F and~II sources in L1688 studied in this paper are shown in Table~\ref{Tab:tab_oph}. 
We report in the following order: 
\begin{itemize}
\item 88 sources observed by \citet{2019ApJ...875L...9W} and classified as members of L1688 by \citet{2020AJ....159..282E}
\item 7 sources observed by \citet{2019ApJ...875L...9W} and classified as candidate members by \citet{2020AJ....159..282E}
\item 3 sources observed by \citet{2019ApJ...875L...9W} in the geographical region of L1688 and with spectral types and magnitudes from \citet{2010ApJS..188...75M,2012ApJ...744..134M,2016AA...592A.126V}
\item 7 VLMS and BDs sources observed either as part of this study or in \citet{2016AA...593A.111T} and classified as members of L1688 by \citet{2020AJ....159..282E}
\end{itemize}

\begin{table*}
\caption{Properties of L1688 objects}
\label{Tab:tab_oph}
\tiny
\begin{tabular}{ccccccccccccccc}
\hline \hline
Spitzer & 2MASS & d & SpT & A$_J$ & Teff & Log$_{10}(L_\star)$ & \Mstar & F$_{225GHz}$ & err & M$_{dust}$ & Log$_{10}(L_{acc})$ & Log$_{10}(\dot{M}_{acc})$ \\
 &      &(pc)&&& (K)&   ($L_\odot$) &(M$_\odot$)& (mJy) & (mJy) & (M$_\oplus$) & ($L_\odot/yr$) &($M_\odot/yr$)\\
\hline
J162459.8-245601 & 16245974-2456008 & 131.99 & M3.5 & 1.08 & 3342 & -0.59 & 0.26 & 2.91 & 0.31 & 1.5 & -- & -- \\
J162502.1-245932 & 16250208-2459323 & 142.04 & M3 & 0.92 & 3415 & -0.5 & 0.29 & 13.1 & 0.31 & 7.9 & -- & --  \\
J162506.2-244657 & 16250623-2446570 & 146.24 & M5 & 1.45 & 3125 & -1.26 & 0.16 & 3.02 & 0.12 & 1.9 & -- & --  \\
J162522.1-241751 & 16252214-2417505 & 143.37 & M4 & 2.08 & 3270 & -0.92 & 0.23 & $\le$0.55 & -- & 0.34 & -- & --  \\
J162524.3-242944 & 16252434-2429442 & 133.59 & M4 & 0.63 & 3270 & -0.63 & 0.22 & 12.4 & 0.15 & 6.6 & -- & -- \\
 \dots  &  \dots  &  \dots  &  \dots  & \dots & \dots  & \dots & \dots & \dots & \dots & \dots & \dots & \dots \\
\hline
\end{tabular}
\tablefoot{({\tiny Table B1 is only available in electronic form
at the CDS via anonymous ftp to cdsarc.u-strasbg.fr (130.79.128.5)
or via {\tt http://cdsweb.u-strasbg.fr/cgi-bin/qcat?J/A+A/}})}
\end{table*}

\section{Lupus data}
\label{App:tab_lupus}

Parameters of the Class~F and~II sources in Lupus studied in this paper are shown in Table~\ref{Tab:tab_lup}.

\begin{table*}
\caption{Properties of Lupus objects}
\label{Tab:tab_lup}
\tiny
\begin{tabular}{cccccccccccc}
\hline \hline
2MASS & d & Teff & Log$_{10}(L_\star)$ & \Mstar & F$_{225GHz}$ & err & M$_{dust}$ & Log$_{10}(L_{acc})$ & Log$_{10}(\dot{M}_{acc})$ \\
      &(pc)& (K)& ($L_\odot$)&(M$_\odot$)& (mJy) & (mJy) & (M$_\oplus$) &($L_\odot/yr$) &($M_\odot/yr$)\\

\hline
J15392776-3446171 & 155.29 & 4060 & -0.05 & 0.7 & 29.94 & 0.2 & 21.6 & $\le$ -2.57 & $\le$-9.54 \\
J15392828-3446180 & 157.34 & 3415 & -0.66 & 0.29 & 6.42 & 0.18 & 4.7 &  -1.76 & -8.5 \\
J15450887-3417333 & 154.96 & 3060 & -1.22 & 0.14 & 20.7 & 0.18 & 14.9 &  -1.77 & -8.39 \\
J15451286-3417305 & 154.19 & 4900 & 0.73 & 2.15 & 66.38 & 0.2 & 47.3 & $\le$ -1.18 & $\le$-8.41 \\
J15451741-3418283 & 154.55 & 3197 & -1.05 & 0.2 & 8.05 & 0.15 & 5.7 &  -2.77 & -9.49 \\
 \dots  &  \dots  &  \dots  &  \dots  & \dots & \dots  & \dots & \dots & \dots & \dots \\
\hline
\end{tabular}
\tablefoot{({\tiny Table C1 is only available in electronic form
at the CDS via anonymous ftp to cdsarc.u-strasbg.fr (130.79.128.5)
or via {\tt http://cdsweb.u-strasbg.fr/cgi-bin/qcat?J/A+A/}})}
\end{table*}

\section{Upper Scorpius data}
\label{App:tab_usc}

Parameters of the Class~F and~II sources in Upper Scorpius studied in this paper are shown in Table~\ref{Tab:tab_usc}.

\begin{table*}
\caption{Properties of Upper Scorpius objects}
\label{Tab:tab_usc}
\tiny
\begin{tabular}{cccccccccccc}
\hline \hline
2MASS & d & Teff & Log$_{10}(L_\star)$ & \Mstar & F$_{345GHz}$ & err & M$_{dust}$ & Log$_{10}(L_{acc})$ & Log$_{10}(\dot{M}_{acc})$ \\
      &(pc)& (K)&  (L$_\odot$)        &(M$_\odot$)& (mJy) & (mJy) & (M$_\oplus$) &(L$_\odot$/yr) &($M_\odot/yr$)\\
 \hline
J15354856-2958551 & 145.0 & 3236 & -0.6 & 0.2 & 1.92 & 0.15 & 0.48 & -2.77 & -9.45 \\
J15514032-2146103 & 142.19 & 3236 & -1.33 & 0.2 & 0.76 & 0.16 & 0.18 & -3.48 & -10.3 \\
J15521088-2125372 & 167.64 & 3236 & -1.68 & 0.19 & $\le$0.45 & -- & 0.15 &  -- & --   \\
J15530132-2114135 & 146.35 & 3236 & -1.19 & 0.21 & 5.78 & 0.14 & 1.5 & -3.01 & -9.82 \\
J15534211-2049282 & 135.63 & 3311 & -0.9 & 0.25 & 2.93 & 0.29 & 0.64 & -2.62 & -9.44 \\
 \dots  &  \dots  &  \dots  &  \dots  & \dots & \dots  & \dots & \dots & \dots & \dots \\
\hline
\end{tabular}
\tablefoot{({\tiny Table C1 is only available in electronic form
at the CDS via anonymous ftp to cdsarc.u-strasbg.fr (130.79.128.5)
or via {\tt http://cdsweb.u-strasbg.fr/cgi-bin/qcat?J/A+A/}})}
\end{table*}

\section{Chamaeleon~I data}
\label{App:tab_cha}

Parameters of the Class~F and~II sources in Chamaeleon~I studied in this paper are shown in Table~\ref{Tab:tab_cha}.

\begin{table*}
\caption{Properties of Chamaeleon I objects}
\label{Tab:tab_cha}
\tiny
\begin{tabular}{cccccccccccc}
\hline \hline
2MASS & d & Teff & Log$_{10}(L_\star)$ & \Mstar & F$_{345GHz}$ & err & M$_{dust}$ & Log$_{10}(L_{acc})$ & Log$_{10}(\dot{M}_{acc})$ \\
      &(pc)& (K)&  (L$_\odot$)        &(M$_\odot$)& (mJy) & (mJy) & (M$_\oplus$) &(L$_\odot$/yr) &($M_\odot/yr$)\\
      \hline
J10533978-7712338 & 191.81 & 3560 & -1.7 & 0.33 & 4.6 & 0.79 & 2.0 & -4.4 & -11.72 \\
J10555973-7724399 & 185.08 & 4060 & -0.74 & 0.81 & 34.1 & 1.32 & 14.0 & -1.12 & -8.43 \\
J10561638-7630530 & 196.48 & 2935 & -1.1 & 0.14 & 3.99 & 0.16 & 1.8 & -4.37 & -10.78 \\
J10574219-7659356 & 190.0 & 3415 & -0.28 & 0.32 & 9.12 & 0.83 & 4.0 & -1.83 & -8.35 \\
J10580597-7711501 & 186.57 & 3060 & -2.0 & 0.11 & 2.68 & 0.16 & 1.1 & -4.94 & -11.68 \\
 \dots  &  \dots  &  \dots  &  \dots  & \dots &  \dots & \dots & \dots & \dots & \dots \\
\hline
\end{tabular}
\tablefoot{({\tiny Table E1 is only available in electronic form
at the CDS via anonymous ftp to cdsarc.u-strasbg.fr (130.79.128.5)
or via {\tt http://cdsweb.u-strasbg.fr/cgi-bin/qcat?J/A+A/}})}
\end{table*}

\section{Corona Australis data}
\label{App:tab_cra}

Parameters of the Class~F and~II sources in Corona Australis studied in this paper are shown in Table~\ref{Tab:tab_cra}.

\begin{table*}
\caption{Properties of Corona Australis objects}
\label{Tab:tab_cra}
\tiny
\begin{tabular}{ccccccccc}
\hline \hline
2MASS & d & Teff & Log$_{10}(L_\star)$ & \Mstar & F$_{225GHz}$ & err & M$_{dust}$ \\
      &(pc)& (K)&  (L$_\odot$)        &(M$_\odot$)& (mJy) & (mJy) & (M$_\oplus$)\\
\hline
J18563974-3707205 & 159.19 & 2860 & -0.9 & 0.1 & $\le$0.3 & -- & 0.23 \\
J19005974-3647109 & 143.58 & 3190 & -1.83 & 0.16 & 0.65 & 0.11 & 0.40 \\
J19011629-3656282 & 156.05 & 2980 & -1.51 & 0.09 & 1.37 & 0.24 & 1.0 \\
J19011893-3658282 & 149.32 & 3560 & -0.61 & 0.38 & 2.77 & 0.26 & 1.8 \\
J19013232-3658030 & 160.89 & 3300 & -0.78 & 0.24 & $\le$0.24 & -- & 0.18 \\
 \dots  &  \dots  &  \dots  &  \dots  & \dots & \dots  & \dots & \dots  \\
\hline
\end{tabular}
\tablefoot{({\tiny Table F1 is only available in electronic form
at the CDS via anonymous ftp to cdsarc.u-strasbg.fr (130.79.128.5)
or via {\tt http://cdsweb.u-strasbg.fr/cgi-bin/qcat?J/A+A/}})}
\end{table*}

\section{Taurus data}
\label{App:tab_tau}

Parameters of the Class~F and~II sources in Taurus studied in this paper are shown in Table~\ref{Tab:tab_tau}.

\begin{table*}
\caption{Properties of Taurus objects}
\label{Tab:tab_tau}
\tiny
\begin{tabular}{ccccccccc}
\hline \hline
2MASS & d & Teff & Log$_{10}(L_\star)$ & \Mstar & F$_{225GHz}$ & err & M$_{dust}$ \\
      &(pc)& (K)&   (L$_\odot$)       &(M$_\odot$)& (mJy) & (mJy) & (M$_\oplus$)\\
\hline
J04183158+2816585 & 128.39 & 3415 & -0.43 & 0.29 & $\le$0.36 & -- & 0.2 \\
J04183158+2816585 & 128.39 & 2990 & -1.21 & 0.11 & 0.6 & 0.12 & 0.3 \\
J04220217+2657304 & 160.31 & 3350 & -0.65 & 0.26 & $\le$0.41 & -- & 0.3 \\
J04220217+2657304 & 160.31 & 3800 & -0.7 & 0.6 & 1.9 & 0.14 & 1.5 \\
J04302961+2426450 & 130.11 & 3800 & 0.12 & 0.56 & 5.6 & 0.12 & 2.8 \\
 \dots  &  \dots  &  \dots  &  \dots  & \dots & \dots & \dots & \dots \\
\hline
\end{tabular}
\tablefoot{({\tiny Table G1 is only available in electronic form
at the CDS via anonymous ftp to cdsarc.u-strasbg.fr (130.79.128.5)
or via {\tt http://cdsweb.u-strasbg.fr/cgi-bin/qcat?J/A+A/}})}
\end{table*}

\section{\Mdust\ vs. \Mstar, and \Macc\ vs. \Mstar\ regression}
\label{App:fits}   

In this appendix we show the results of applying the \citet{2007ApJ...665.1489K} method to the samples analyzed in 
this paper. We used the python implementation of the {\tt linmix} code\footnote{\tt https://linmix.readthedocs.io/en/latest/ and https://github.com/jmeyers314/linmix}. In all cases we included
the non-detections in the dataset, properly accounting for
censored data in the computations.

\begin{figure*}
   \centering
       \includegraphics[width=18cm]{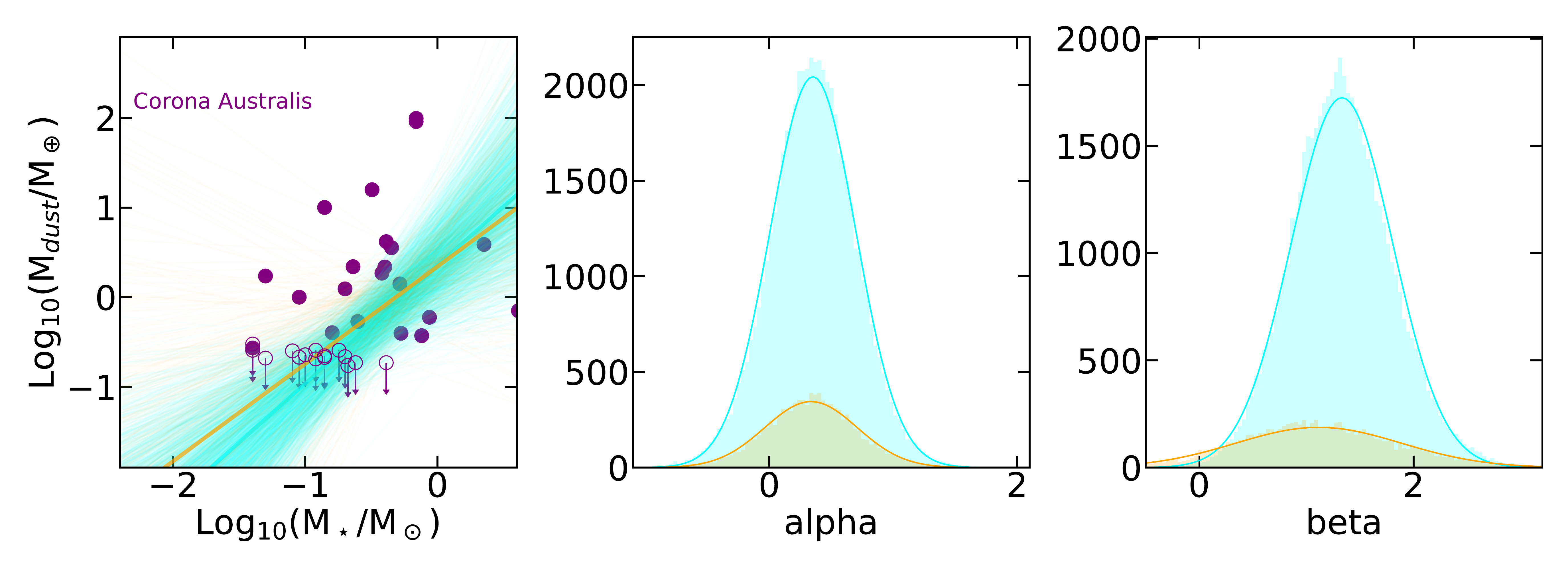}
       \includegraphics[width=18cm]{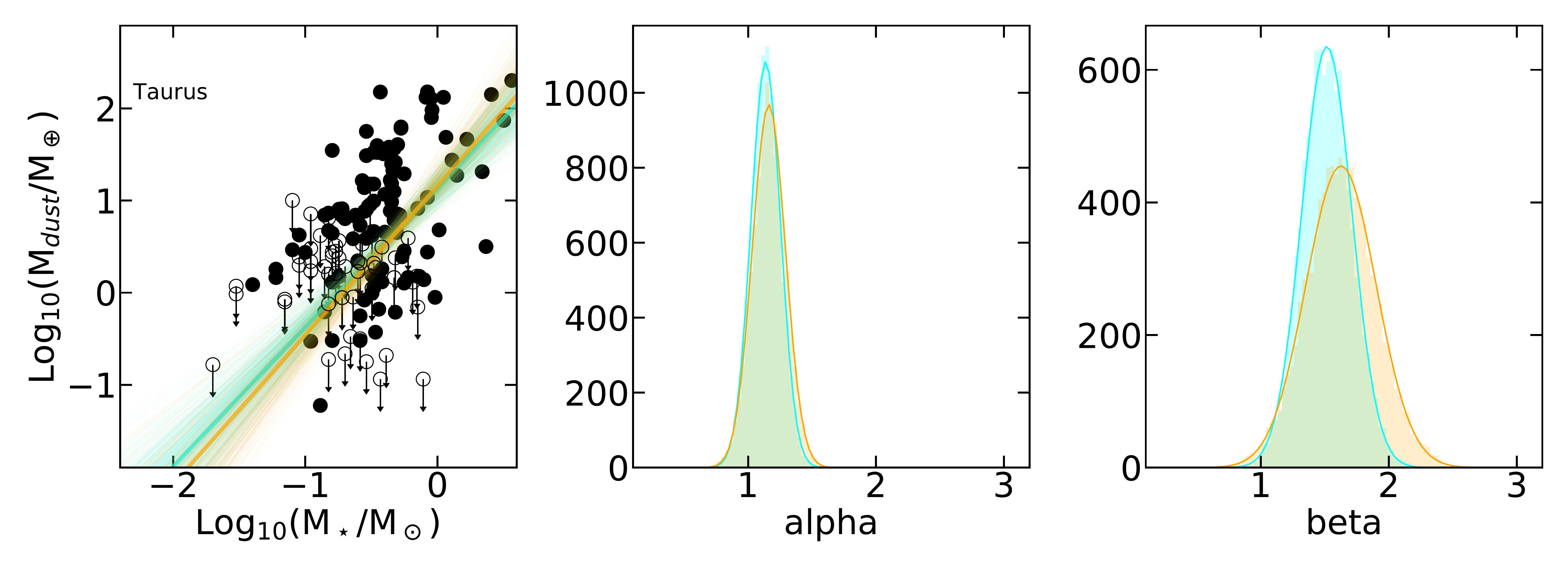}         \includegraphics[width=18cm]{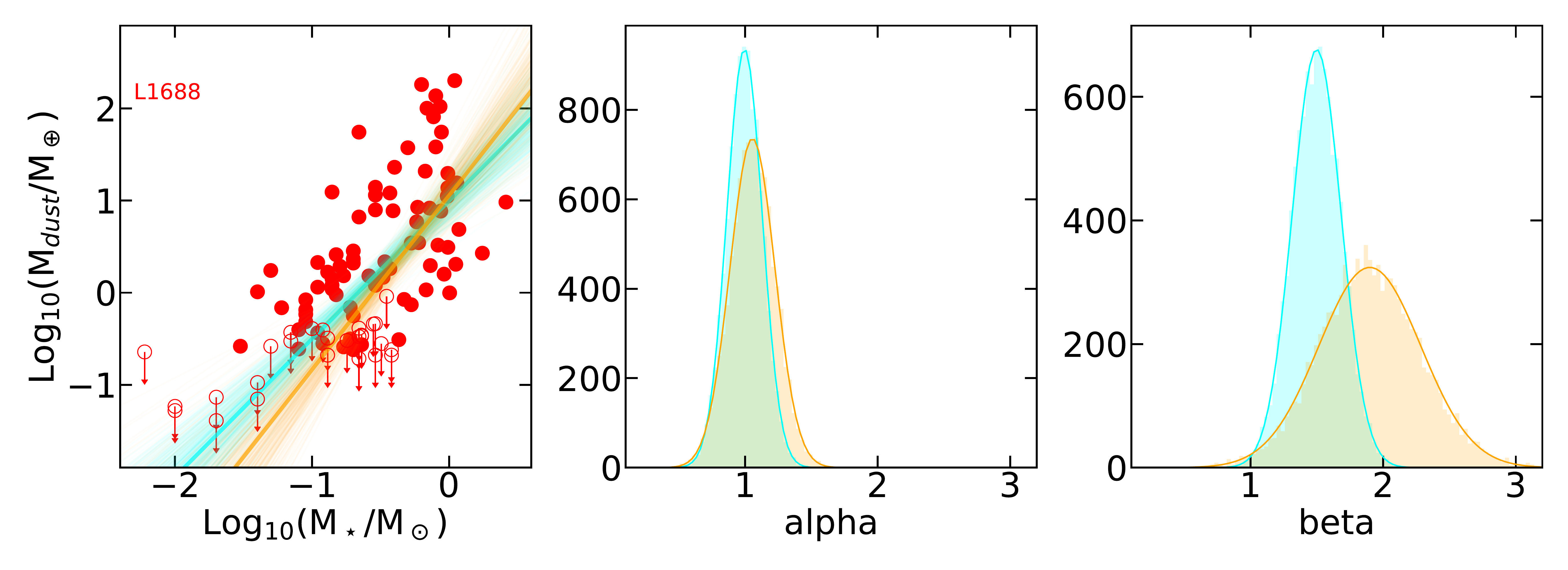}       \caption{Power-law fits to the \Mdust\ vs. \Mstar\ relationships. From top to bottom: Results for Corona Australis, Taurus, and L1688. From left to right: Plot of the dataset and fit results, probability distributions for the $\alpha$ parameter in Eq.~\ref{Eq:fit}, and for the $\beta$ parameter in Eq.~\ref{Eq:fit}. Cyan is when all objects are included, orange when  objects with \Mstar$< 0.15$ \Msun\ are excluded. 
              }
         \label{Fig:Mdust_Mstar_fits_cra_tau_oph}
\end{figure*}

\begin{figure*}
   \centering
\includegraphics[width=18cm]{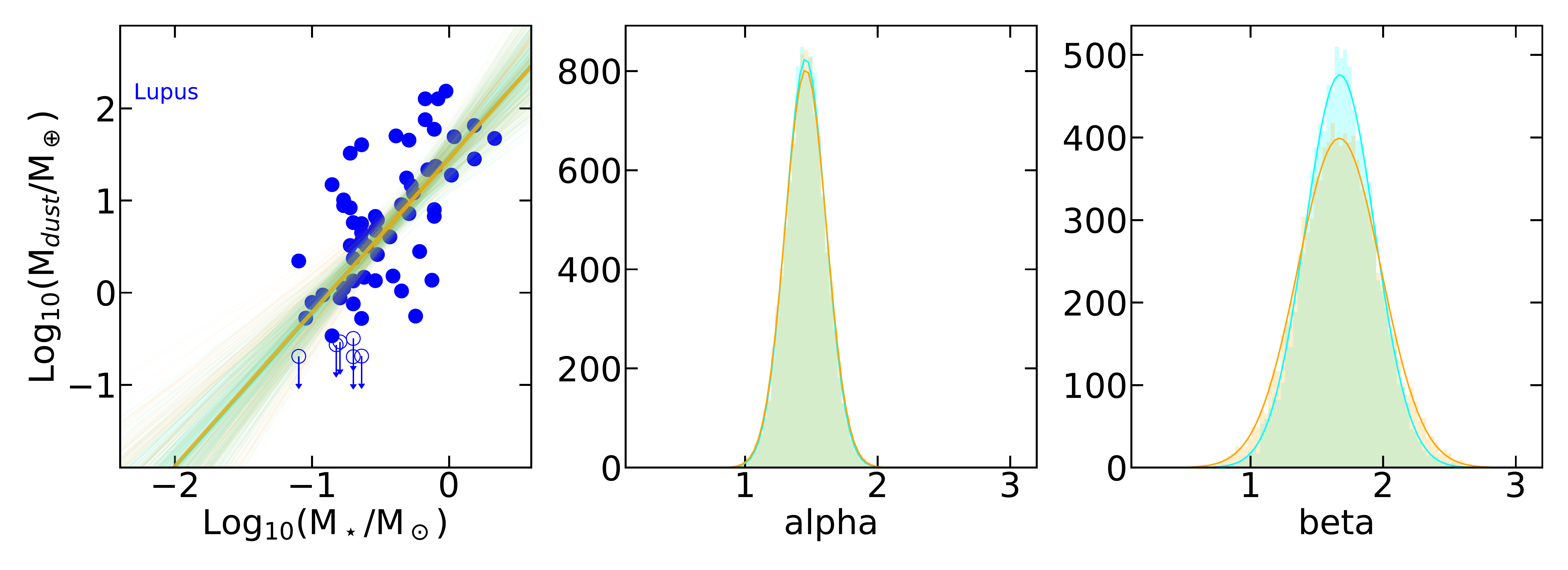} 
\includegraphics[width=18cm]{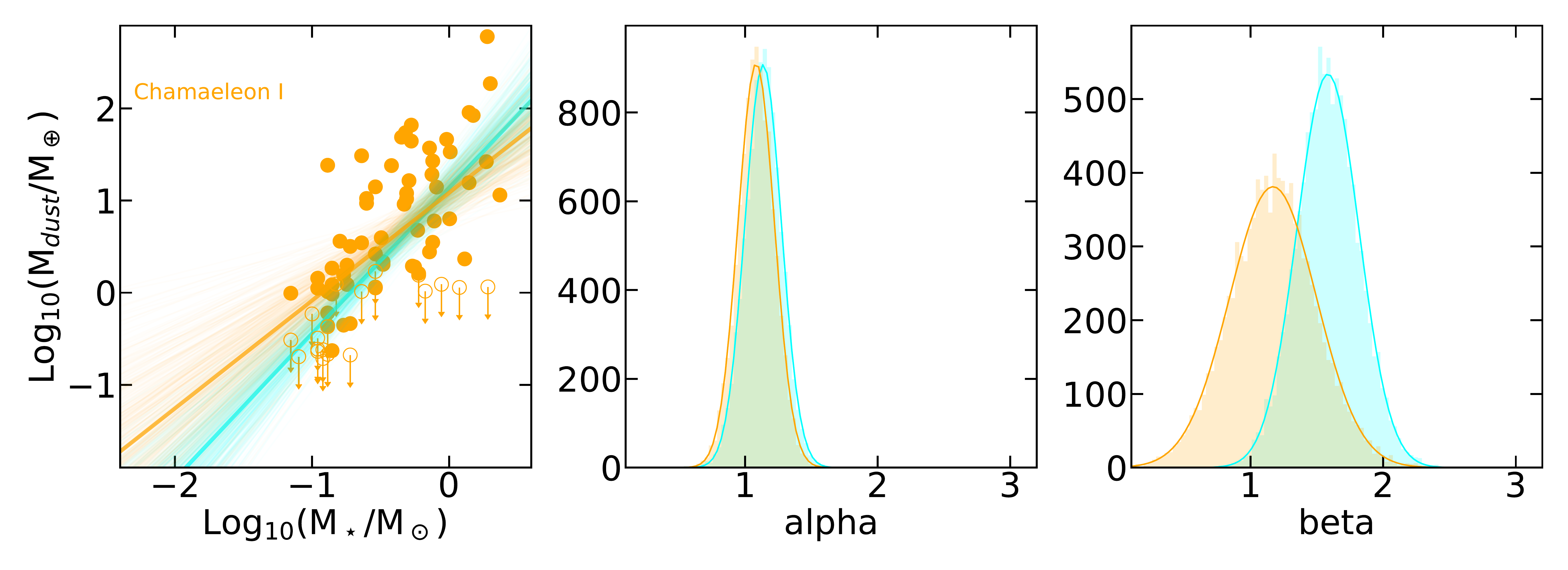}
\includegraphics[width=18cm]{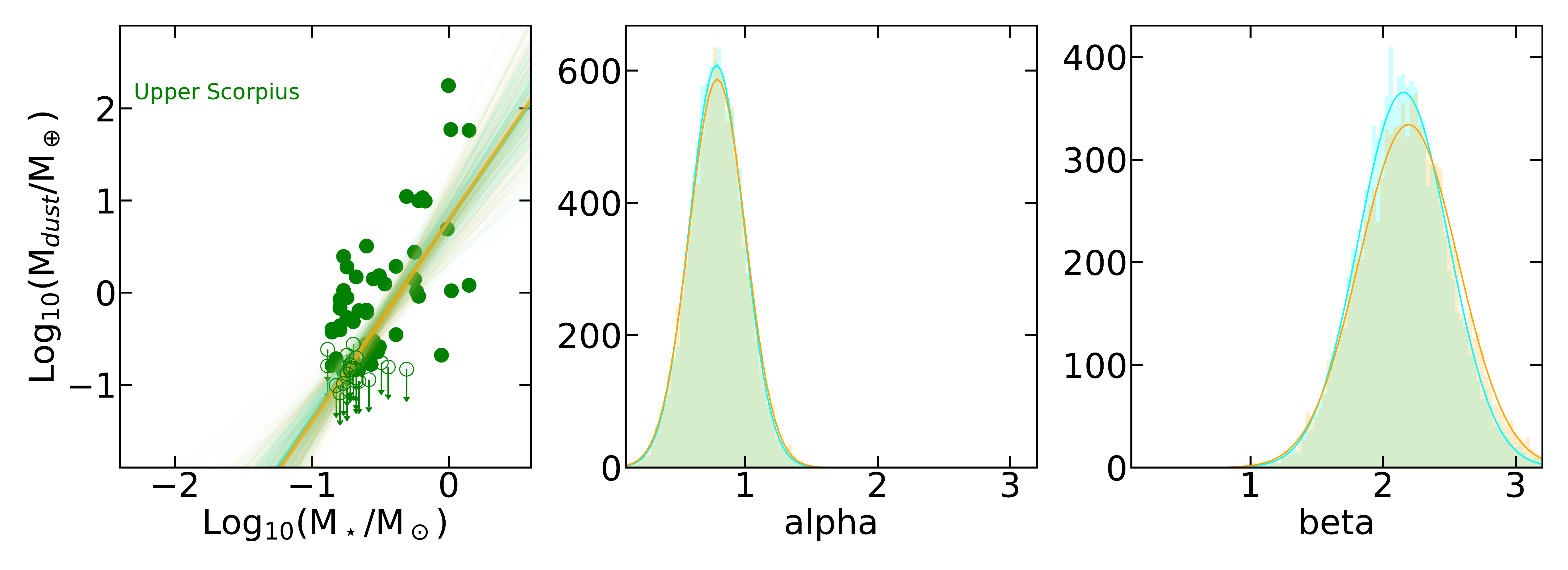}
      \caption{
       Power-law fits to the \Mdust\ vs. \Mstar\ relationships. From top to bottom: Results for  Lupus,
       Chamaeleon~I, and Upper Scorpius. From left to right: Plot of the dataset and fit results, probability distributions for the $\alpha$ parameter in Eq.~\ref{Eq:fit}, and for the $\beta$ parameter in Eq.~\ref{Eq:fit}. Cyan is when all objects are included, orange when  objects with \Mstar$< 0.15$ \Msun\ are excluded.
              }
         \label{Fig:Mdust_Mstar_fits_lup_cha_usc}
\end{figure*}

In Figs.~\ref{Fig:Mdust_Mstar_fits_cra_tau_oph} and~\ref{Fig:Mdust_Mstar_fits_lup_cha_usc} we show the results of fitting
the following relation:
\begin{equation}
    Log_{10}(M_{dust}/M_\oplus) = \alpha + \beta * Log_{10}(M_{star}/M_\odot)
\label{Eq:fit}
\end{equation}
to the Corona Australis, Taurus, and L1688 (Fig.~\ref{Fig:Mdust_Mstar_fits_cra_tau_oph}, from top to bottom), and Lupus, Chamaeleon~I and Upper Scorpius (Fig.~\ref{Fig:Mdust_Mstar_fits_lup_cha_usc}, from top to bottom) samples. In all figures the left panel shows the data and the resulting fit including the full population (cyan) and only the subsample with \Mstar$\ge 0.15$\Msun\ (orange).

\begin{table*}
    \caption{Results of the power-law fits to the \Mdust\ vs. \Mstar\ relations. The value of $\delta$ is a measure of the dispersion from the fit (see {\tt linmix} documentation).}
    \centering
    {
    \begin{tabular}{|l|ccc|ccc|}
         \hline
         \Mdust-\Mstar & \multicolumn{6}{c|}{Log$_{10}$(M$_{dust}$/M$_\oplus$) = $\alpha$ + $\beta$ * Log$_{10}$(M$_{star}$/M$_\odot$)}
         \\
         & $\alpha$ & $\beta$ & $\delta$ & $\alpha$ & $\beta$ & $\delta$\\
         Region &\multicolumn{3}{c|}{full sample}&\multicolumn{3}{c|}{\Mstar$\ge 0.15$\Msun}\\
         \hline
Corona Australis &  0.4$\pm$0.4 & 1.3$\pm$0.5 & 1.1$\pm$0.7 
                &  0.3$\pm$0.4 & 1.1$\pm$0.8 & 1.1$\pm$0.8 \\
         Taurus &  1.1$\pm$0.1 & 1.5$\pm$0.2 & 0.8$\pm$0.3 
                &  1.2$\pm$0.1 & 1.6$\pm$0.3 & 0.9$\pm$0.3 \\
         L1688 &  1.0$\pm$0.1 & 1.5$\pm$0.2 & 0.8$\pm$0.3 
               &  1.1$\pm$0.2 & 1.9$\pm$0.4 & 0.9$\pm$0.4 \\
         Lupus &  1.4$\pm$0.2 & 1.7$\pm$0.3 & 0.7$\pm$0.3 
               &  1.5$\pm$0.2 & 1.7$\pm$0.3 & 0.7$\pm$0.3 \\
Chamaeleon~I   & 1.1$\pm$0.2 & 1.6$\pm$0.3 & 0.7$\pm$0.4 
               & 1.1$\pm$0.2 & 1.2$\pm$0.3 & 0.7$\pm$0.4\\
Upper Scorpius &  0.8$\pm$0.2 & 2.2$\pm$0.3 & 0.7$\pm$0.3 
               &  0.8$\pm$0.2 & 2.2$\pm$0.4 & 0.7$\pm$0.4 \\
         \hline
         \Macc-\Mstar & \multicolumn{6}{c|}{Log$_{10}$($\dot{M}_{acc}$/(M$_\odot$/yr)) = $\alpha$ + $\beta$ * Log$_{10}$(M$_{star}$/M$_\odot$)}\\
         & $\alpha$ & $\beta$ & $\delta$ & $\alpha$ & $\beta$ & $\delta$\\
         Region &\multicolumn{3}{c|}{full sample}&\multicolumn{3}{c|}{\Mstar$\ge 0.15$\Msun}\\
         \hline
Corona Australis & -- & -- & --
               & -- & -- & -- \\
         Taurus & -- & -- & -- 
               & -- & -- & -- \\
         L1688 &  $-$8.3$\pm$0.3 & 1.8$\pm$0.5 & 1.2$\pm$0.7
               & $-$8.3$\pm$0.3 & 1.7$\pm$0.7 & 1.2$\pm$0.7\\
         Lupus & $-$8.5$\pm$0.2 & 1.6$\pm$0.3 & 0.7$\pm$0.3
               & $-$8.6$\pm$0.2 & 1.2$\pm$0.4 & 0.7$\pm$0.4\\
Chamaeleon~I   & $-$8.1$\pm$0.2 & 2.3$\pm$0.3 & 1.0$\pm$0.4
               & $-$8.2$\pm$0.2 & 1.3$\pm$0.4 & 1.0$\pm$0.5\\ 
Upper Scorpius & $-$9.0$\pm$0.5 & 1.5$\pm$0.8 & 1.3$\pm$0.8
               & $-$9.0$\pm$0.5 & 1.5$\pm$0.9 & 1.4$\pm$0.8\\
        \hline
        \Macc-\Mdisk & \multicolumn{6}{c|}{Log$_{10}$($\dot{M}_{acc}$/(M$_\odot$/yr)) = $\alpha$ + $\beta$ * Log$_{10}$(M$_{disk}$/M$_\odot$)}\\
         & $\alpha$ & $\beta$ & $\delta$ & $\alpha$ & $\beta$ & $\delta$\\
         Region &\multicolumn{3}{c|}{full sample}&\multicolumn{3}{c|}{\Mstar$\ge 0.15$\Msun}\\
         \hline
Corona Australis & -- & -- & --
               & -- & -- & -- \\
         Taurus & -- & -- & -- 
               & -- & -- & -- \\
         L1688 &  $-$6.0$\pm$0.5 & 1.0$\pm$0.2 & 0.9$\pm$0.5
               & $-$6.2$\pm$0.6 & 0.9$\pm$0.2 & 0.9$\pm$0.6\\
         Lupus & $-$7.2$\pm$0.4 & 0.7$\pm$0.2 & 0.7$\pm$0.4
               & $-$7.6$\pm$0.4 & 0.6$\pm$0.2 & 0.7$\pm$0.3\\
Chamaeleon~I   & $-$6.0$\pm$0.6 & 1.1$\pm$0.2 & 1.1$\pm$0.5
               & $-$6.6$\pm$0.6 & 0.7$\pm$0.2 & 1.0$\pm$0.5\\
Upper Scorpius & $-$7.0$\pm$1.3 & 0.8$\pm$0.4 & 1.3$\pm$0.7
               & $-$7.0$\pm$1.4 & 0.8$\pm$0.4 & 1.3$\pm$0.8\\
        \hline
    \end{tabular}
    }
    \label{Tab:fit_mdust_mstar}
\end{table*}

\begin{figure*}
   \centering
       \includegraphics[width=18cm]{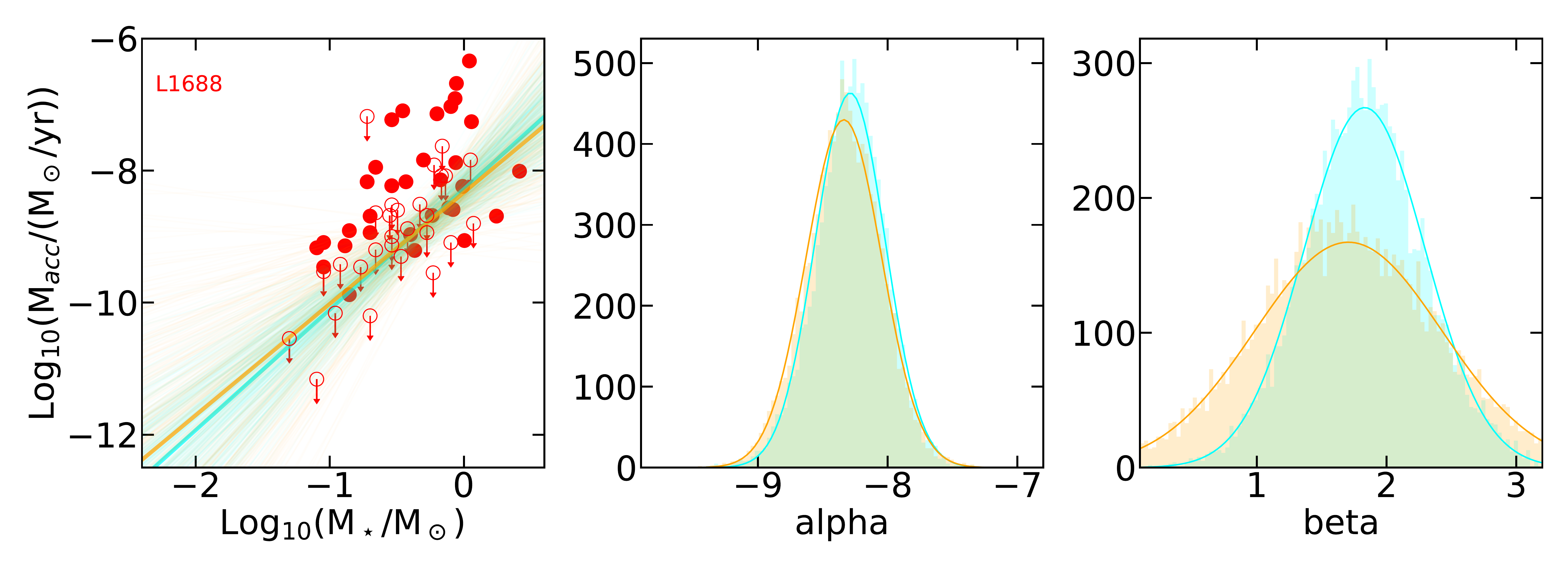} \includegraphics[width=18cm]{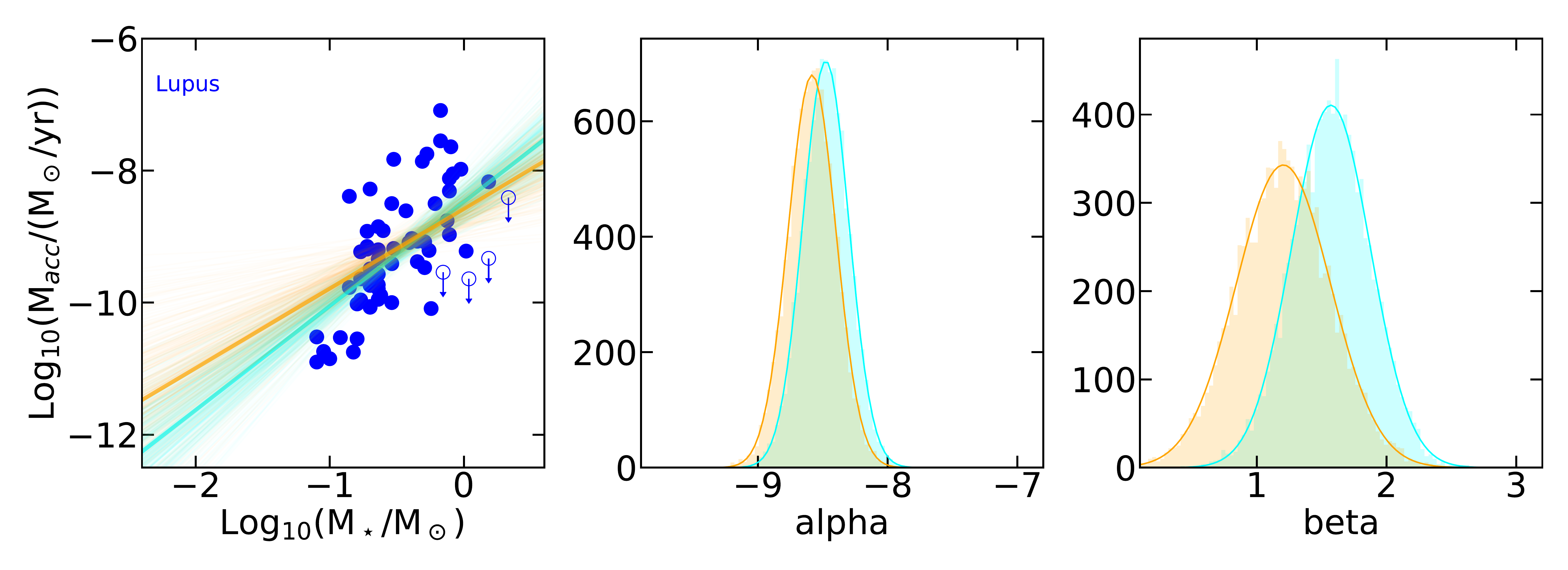} \includegraphics[width=18cm]{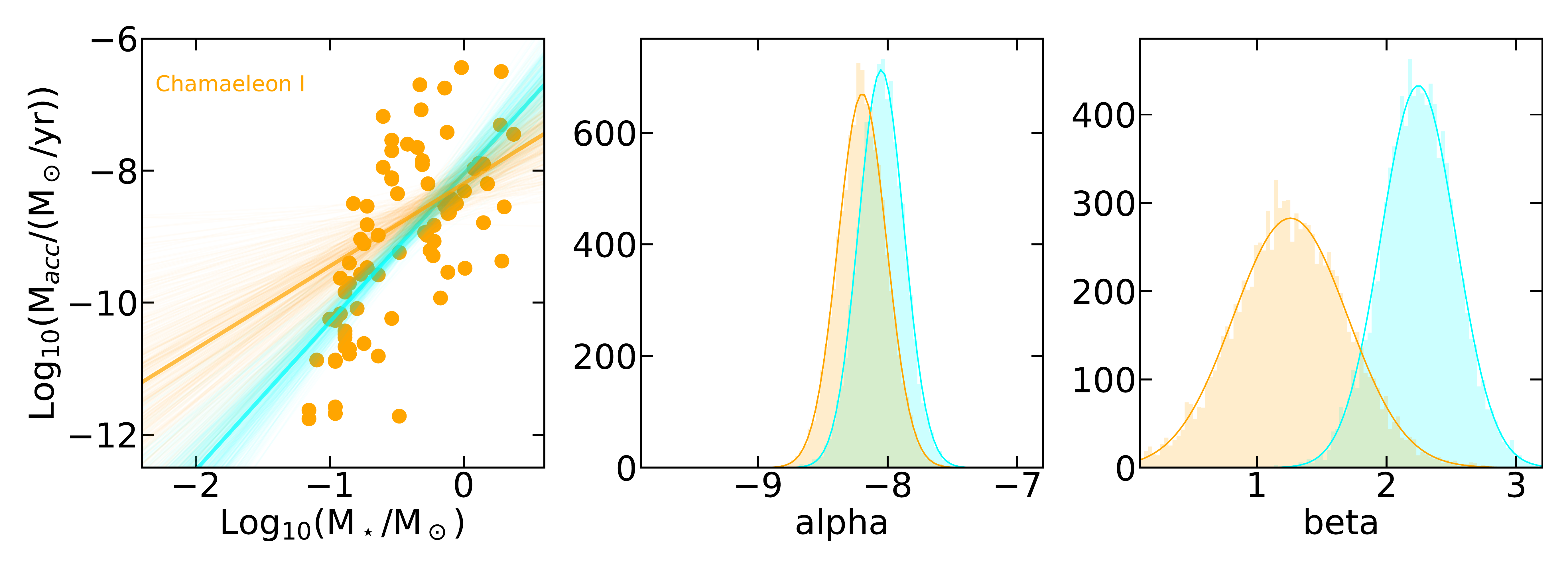} \includegraphics[width=18cm]{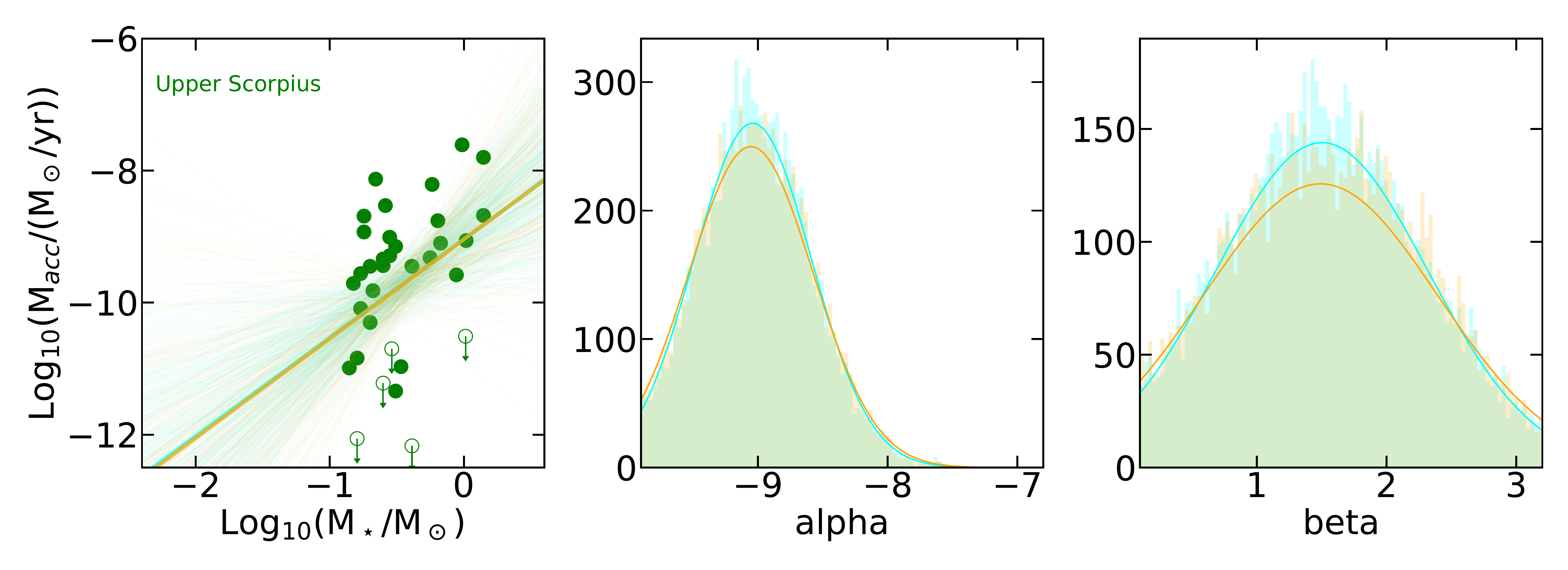} 
      \caption{Power-law fits to the \Macc\ vs. \Mstar\ relationships. From top to bottom: Results for L1688, Lupus, Chamaeleon~I, and Upper Scorpius. From left to right: Plot of the dataset and fit results, probability distributions for the $\alpha$ parameter in Eq.~\ref{Eq:fit_macc}, and for the $\beta$ parameter in Eq.~\ref{Eq:fit_macc}.
              }
         \label{Fig:Macc_Mstar_fits_oph_lup_usc_cha}
\end{figure*}

\begin{figure*}
   \centering
       \includegraphics[width=18cm]{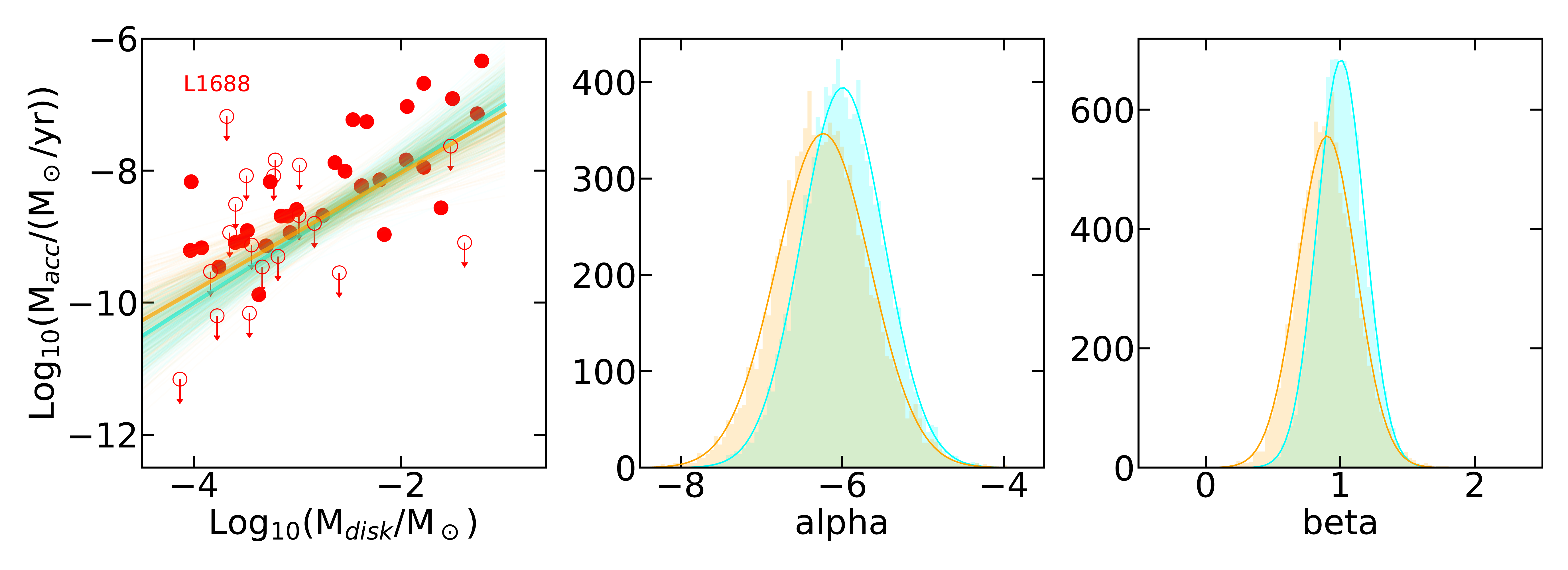} \includegraphics[width=18cm]{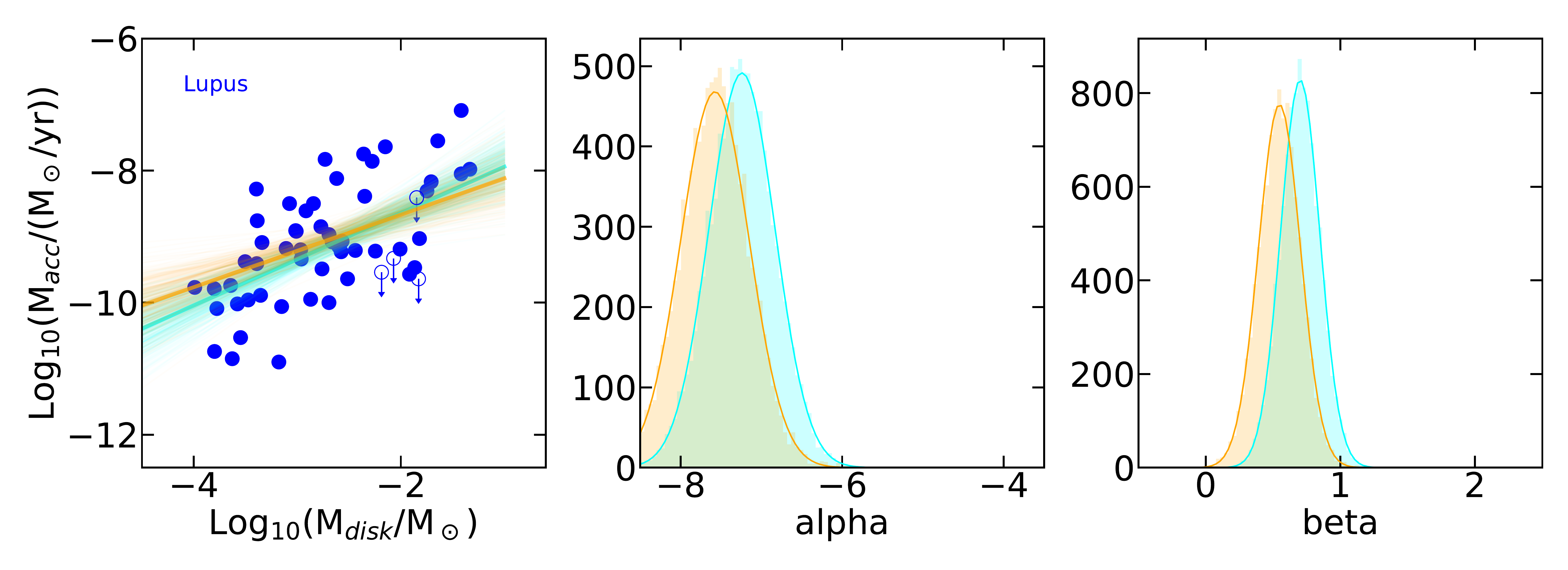} \includegraphics[width=18cm]{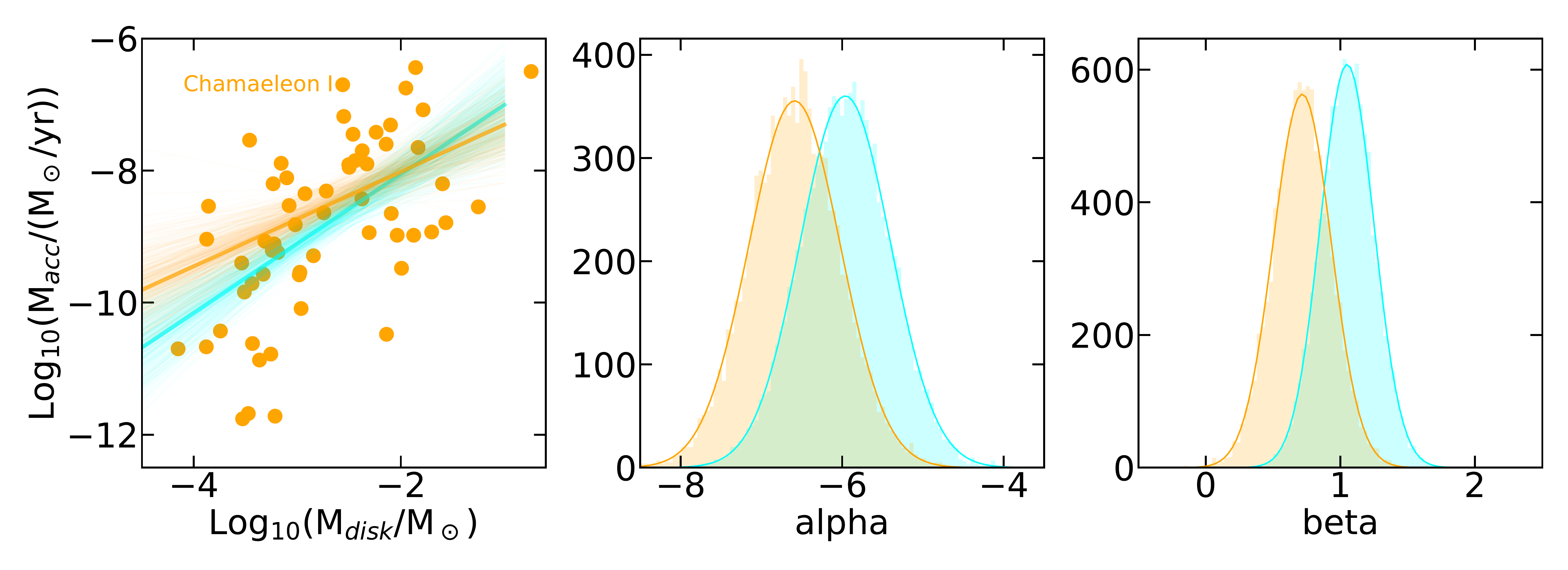} \includegraphics[width=18cm]{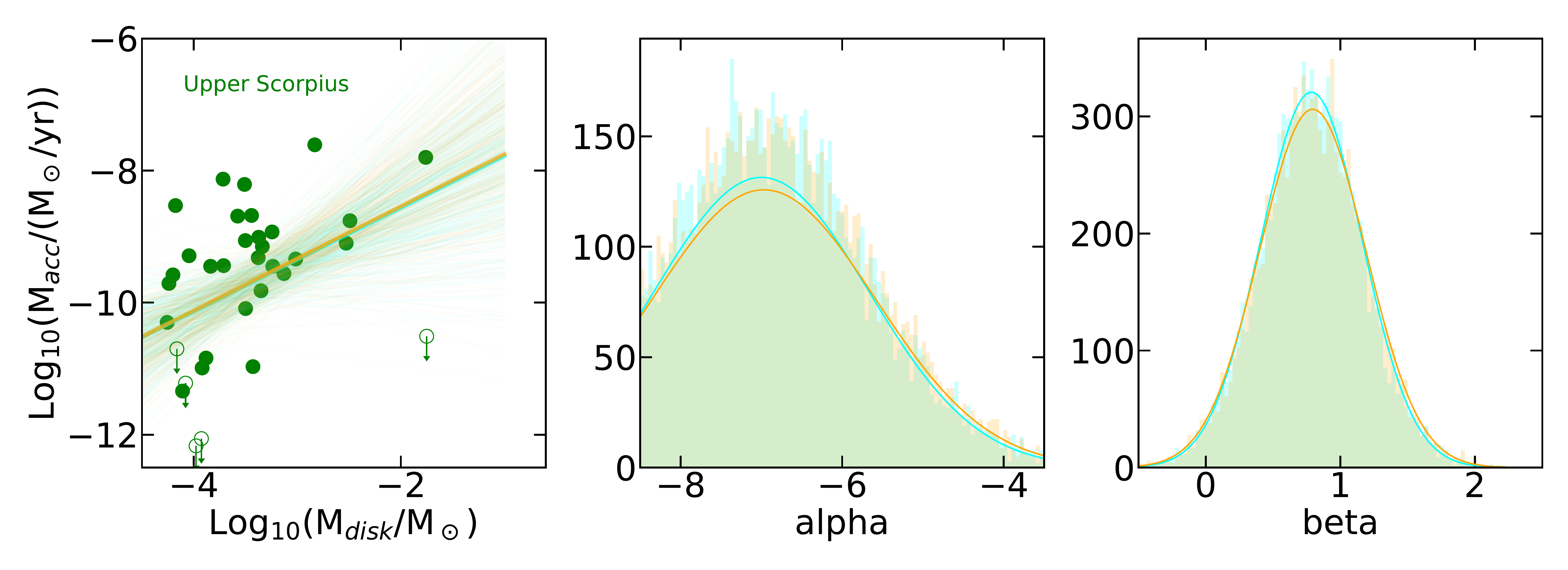} 
      \caption{Power-law fits to the \Macc\ vs. \Mdisk\ relationships. From top to bottom: Results for L1688, Lupus, Chamaeleon~I, and Upper Scorpius. From left to right: Plot of the dataset and fit results, probability distributions for the $\alpha$ parameter in Eq.~\ref{Eq:fit_mamd}, and for the $\beta$ parameter in Eq.~\ref{Eq:fit_mamd}.
              }
         \label{Fig:Macc_Mdisk_fits_oph_lup_usc_cha}
\end{figure*}

In Fig.~\ref{Fig:Macc_Mstar_fits_oph_lup_usc_cha} we show the results of fitting
the following relation:
\begin{equation}
    Log_{10}(\dot{M}_{acc}/(M_\odot/yr)) = \alpha + \beta * Log_{10}(M_{star}/M_\odot)
\label{Eq:fit_macc}
\end{equation}
to the L1688, Lupus, Chamaeleon~I and Upper Scorpius (from top to bottom) samples. In all figures the left panel shows the data and the resulting fit including the full population (cyan) and only the subsample with \Mstar$\ge 0.15$\Msun\ (orange).

In Fig.~\ref{Fig:Macc_Mdisk_fits_oph_lup_usc_cha} we show the results of fitting
the following relation:
\begin{equation}
    Log_{10}(\dot{M}_{acc}/(M_\odot/yr)) = \alpha + \beta * Log_{10}(M_{disk}/M_\odot)
\label{Eq:fit_mamd}
\end{equation}
to the L1688, Lupus, Chamaeleon~I and Upper Scorpius (from top to bottom) samples. In all figures the left panel shows the data and the resulting fit including the full population (cyan) and only the subsample with \Mstar$\ge 0.15$\Msun\ (orange).

The values derived for all regions and all relationships are reported in Table~\ref{Tab:fit_mdust_mstar}. The value of $\delta$ in the table is a measure of the dispersion of the points from the fit (see {\tt linmix} documentation).

\end{appendix}

%
%
\end{document}